\documentclass[aps,prd,twocolumn,groupedaddress,showpacs,eqsecnum,floatfix]{revtex4-1}

\usepackage{amsmath}
\usepackage{amssymb}
\usepackage{graphicx}
\usepackage{bm}
\usepackage{braket}
\usepackage{tensor}
\usepackage{slashed}

\newcommand{\bg}{\bar{\gamma}}

\newcommand{\bh}{\bar{h}}
\newcommand{\barf}{\bar{f}}
\newcommand{\df}{\delta f}
\newcommand{\bpsi}{\bar{\psi}}
\newcommand{\bA}{\bar{A}_0}

\begin{document}

\title{Stability of gravitating charged-scalar solitons in a cavity}

\author{Supakchai Ponglertsakul}
\email{smp12sp@sheffield.ac.uk}

\affiliation{Consortium for Fundamental Physics, School of Mathematics and Statistics,
The University of Sheffield, Hicks Building, Hounsfield Road, Sheffield S3 7RH, United Kingdom}

\author{Elizabeth Winstanley}
\email{e.winstanley@sheffield.ac.uk}

\affiliation{Consortium for Fundamental Physics, School of Mathematics and Statistics,
The University of Sheffield, Hicks Building, Hounsfield Road, Sheffield S3 7RH, United Kingdom}

\affiliation{Department of Physics and Astronomy, University of Canterbury, Private Bag 4800, Christchurch 8140, New Zealand}

\author{Sam R.~Dolan}
\email{s.dolan@sheffield.ac.uk}

\affiliation{Consortium for Fundamental Physics, School of Mathematics and Statistics,
The University of Sheffield, Hicks Building, Hounsfield Road, Sheffield S3 7RH, United Kingdom}

\date{\today}

\begin{abstract}
We present new regular solutions of Einstein charged scalar field theory in a cavity.
The system is enclosed inside a reflecting mirror-like boundary, on which the scalar field vanishes. The mirror is placed at the zero of the scalar field closest to the origin, and  inside this boundary our solutions are regular.
We study the stability of these solitons under linear, spherically symmetric perturbations of the metric, scalar and electromagnetic fields.
If the radius of the mirror is sufficiently large, we present numerical evidence for the stability of the solitons.
For small mirror radius, some of the solitons are unstable.
We discuss the physical interpretation of this instability.
\end{abstract}


\maketitle

\section{Introduction}
\label{sec:intro}
The search for compact-body solutions of General Relativity began in 1915.  A century on, our perspective is that neutron stars and black holes are abundant in our Universe, and that the supernovae which create them are vital in seeding galaxies with heavy elements.

Compact-body solutions divide into two classes: (1) {\it black holes}, causally-nontrivial geometries with spacetime horizons; and (2) {\it solitons}, regular geometries sourced by matter fields. Broadly interpreted, the latter class comprises white dwarfs and neutron stars, as well as exotic hypothetical possibilities, such as quark, preon, electroweak or boson stars \cite{Liebling:2012fv}. Solitons can become black holes in gravitational collapse (cf.~the Chandrasekhar and Tolman-Oppenheimer-Volkoff mass limits). But are there other possibilities?

In certain scenarios, black holes may support ``hair'', in the form of non-trivial matter fields (for a recent review, see e.g.~Ref.~\cite{Volkov:2016ehx}). A hairy black hole may be thought of as a hybrid: a nonlinear superposition of a vacuum black hole and a soliton \cite{Volkov:2016ehx}, or, ``horizons inside lumps'' \cite{Kastor:1992qy}. Like a soliton, a hairy black hole possesses externally-accessible degrees of freedom, yet like a vacuum black hole, it divides spacetime into causally-distinct regions. Many matter models that admit solitonic solutions also admit hairy black hole solutions. In this paper, we seek soliton solutions to  accompany the hairy black hole solutions we identified in Ref.~\cite{Dolan:2015dha}, in the context of Einstein charged scalar field theory \cite{Gundlach:1996vv} in a cavity.

In ${\mathfrak {su}}(2)$ Einstein-Yang-Mills (EYM) theory in four-dimensional, asymptotically flat spacetime, the discovery of solitons \cite{Bartnik:1988am} was followed closely by the discovery of hairy black hole solutions \cite{Volkov:1989fi, Volkov:1990sva, Kunzle:1990, Bizon:1990sr}. However, both the solitons and hairy black holes were soon found to be unstable \cite{Straumann:1989tf, Straumann:1990as, Galtsov:1991du, Volkov:1994dq, Lavrelashvili:1994rp, Volkov:1995np, Hod:2008ir}. Under perturbation, the black holes lose their gauge-field hair and evolve towards a (stable) vacuum black hole solution; the solitons either collapse to form a vacuum black hole or else the gauge field is radiated away to infinity, leaving pure Minkowski spacetime \cite{Zhou:1991nu, Zhou:1992sb, Rinne:2014kka}. This prompts an intriguing question: are there scenarios in which the converse occurs, i.e., in which a vacuum black hole spontaneously evolves towards a hairy configuration which is stable?

It has long been known that, in a Penrose process \cite{Penrose:1971uk}, a vacuum black hole may shed energy and angular momentum (and/or charge) whilst also increasing its horizon area. One such Penrose process is superradiance \cite{Starobinskii}. In the Kerr black hole context, superradiance implies that the low-frequency corotating modes of a bosonic field are scattered with a reflection coefficient of greater than unity (see \cite{Brito:2015oca} for a review). If superradiant modes are trapped in the vicinity of the black hole they suffer repeated amplification, causing exponential growth in the field: a ``black hole bomb'' instability \cite{Press:1972zz}. Various mechanisms for confinement of superradiant modes have been explored, such as a mirror \cite{Press:1972zz, Cardoso:2004nk}, a field mass \cite{Damour:1976kh, Zouros:1979iw, Detweiler:1980uk, Furuhashi:2004jk, Cardoso:2005vk, Dolan:2007mj}, or a spacetime boundary \cite{Cardoso:2004hs, Dold:2015cqa}.

What is the outcome of a black hole bomb instability, in the case of a massive bosonic field bound to a Kerr black hole? One possibility is that the black hole ejects the field in an explosive ``bosenova'' outflow \cite{Yoshino:2012kn, Yoshino:2015nsa}, to return to a near-vacuum configuration. A second possibility is that the black hole evolves towards a hairy configuration which is stable. The latter possibility has been given credence by the recent discovery of an asymptotically-flat family of Kerr-like black holes possessing (complex, massive) scalar-field hair \cite{Herdeiro:2014goa, Herdeiro:2015gia, Herdeiro:2015waa}, and a single (helical) Killing vector \cite{Dias:2011at}. This one-parameter family of solutions to the Einstein-Klein-Gordon equations bifurcates from the Kerr solution \cite{Chodosh:2015nma, Chodosh:2015oma}, and furthermore reduces to ``solitonic'' boson star solutions in a well-defined limit. An important open question is whether such solutions are stable under perturbation.

The timescale for the growth of a massive scalar field instability on Kerr is rather long ($\tau \gtrsim 5.8 \times 10^6 GM / c^3$ \cite{Dolan:2012yt}). This makes it challenging to track the development of the instability from a weak perturbation into the nonlinear regime using a time-domain evolution (though see \cite{Brito:2014wla, Okawa:2014nda, Okawa:2015fsa} and \cite{Witek:2012tr, Zilhao:2015tya} for  scalar and Proca field evolutions, respectively). On the other hand, it is well-known that there is a \emph{charged} version of superradiance \cite{Bekenstein:1973mi}, whereby low-frequency modes of a charged scalar field incident on a spherically-symmetric Reissner-Nordstr\"om black hole are scattered with a reflection coefficient of greater than unity.
Although superradiant bound modes do not form naturally for a charged massive field, due to the electrostatic repulsion that out-competes the gravitational attraction in the superradiant regime, the charged black hole bomb can nevertheless be triggered with two mechanisms: either by placing the charged black hole in a cavity (i.e.~confining the bosonic field within a reflecting mirror) \cite{Herdeiro:2013pia, Degollado:2013bha, Hod:2013fvl, Hod:2016nsr}; or by embedding the charged black hole within an asymptotically anti-de Sitter (adS) spacetime \cite{Hawking:1999dp, Uchikata:2011zz, Dold:2015cqa, Li:2016kws}.
These systems are more tractable for nonlinear studies than the superradiant instability on Kerr, not only because of the simplification afforded by spherical symmetry, but also because the timescales for the development of the charged black hole bomb instability are typically shorter than those in the Kerr case.

Two recent studies of the nonlinear development have shed new light on the ultimate fate of the charged black hole bomb instability. In the cavity scenario, Sanchis-Gual {\it et al.}~\cite{Sanchis-Gual:2015lje} evolved the Einstein-Maxwell-Klein-Gordon equations in the spherically-symmetric sector, and demonstrated that a Reissner-Nordstr\"om black hole in electrovacuum, after weak perturbation, can develop into a hairy configuration \cite{Dolan:2015dha} in which some, but not all, of the charge has transferred from the black hole into the scalar field. For low field charge, the approach to the final state is smooth, whereas for high charge an overshoot triggers an explosive bosenova phenomenon.

In the adS scenario in four dimensions, Bosch {\it et al.}~\cite{Bosch:2016vcp} demonstrated a compatible result: a Reissner-Nordstr\"om-adS black hole, under generic perturbation by a charged scalar field, will develop into a stable hairy configuration, by transferring mass and charge into the surrounding field.

Previously, in Ref.~\cite{Dolan:2015dha} we constructed hairy black hole solutions for Einstein charged scalar field theory \cite{Gundlach:1996vv} in a cavity. By applying a first-order perturbation analysis, we argued that the configuration without zeros in the scalar field between the horizon and mirror would be stable; and that higher overtones with nodes would be unstable. This result was then borne out by the dynamical investigation of Sanchis-Gual {\it et al.}~\cite{Sanchis-Gual:2015lje}. Here, we repeat the analysis for the solitonic case, in anticipation that future time-domain studies will, once again, test our inferences on stability.

The outline of this paper is as follows. In Sec.~\ref{sec:equil} we introduce Einstein charged scalar field theory and the field equations for spherically-symmetric configurations.
Numerical solutions of the field equations representing static charged-scalar solitons in a cavity are presented in Sec.~\ref{sec:solitons}.
The stability of these charged-scalar solitons under linear, spherically-symmetric perturbations of the massless scalar field, electromagnetic field and metric is explored in Sec.~\ref{sec:stab}.
Discussion of the physical interpretation of our stability results and our conclusions can be found in Sec.~\ref{sec:conc}.

\section{Einstein charged scalar field theory}
\label{sec:equil}

We consider the following action, describing Einstein-charged scalar field theory:
\begin{align}
S = \int \sqrt{-g}\left[\frac{R}{2}-\frac{1}{4}F_{\mu \nu  }F^{\mu \nu  } -\frac{1}{2} g^{\mu \nu } D^\ast_{(\mu } \Phi^\ast D^{}_{\nu )} \Phi\right]d^4 x,
\label{eq:action}
\end{align}
where $g$ is the determinant of the metric, $R$ is the Ricci scalar and round brackets denote symmetrization,
$X_{(\mu \nu )} = \frac{1}{2} \left(X_{\mu \nu } + X_{\nu \mu }\right)$ for a tensor field $X_{\mu \nu }$.
Throughout this paper we use a positive spacetime signature $+2$, and units in which $8\pi G = c=1$.
The massless scalar field $\Phi $ is complex, and $\Phi ^{\ast }$ denotes the complex conjugate of $\Phi $.
The electromagnetic field strength $F_{\mu \nu }$ is given by
\begin{equation}
F_{\mu \nu } = \nabla_\mu A_{\nu } - \nabla_\nu A_{\mu} ,
\end{equation}
where $A_{\mu }$ is the electromagnetic potential.
In (\ref{eq:action}), we have introduced
\begin{equation}
D_{\mu } = \nabla_{\mu } - i q A_{\mu } ,
\end{equation}
where $\nabla _{\mu }$ is the covariant derivative and $q$ is the charge of the scalar field $\Phi $.

Varying the action (\ref{eq:action}) yields the field equations
\begin{subequations}
\label{eq:field}
\begin{align}
G_{\mu \nu } &= T_{\mu \nu },
\label{eq:EFE}\\
\nabla_\mu F^{\mu \nu } &= J^\nu ,
\label{eq:MW}\\
D_{\mu }D^{\mu }\Phi &= 0 .
\label{eq:KG}
\end{align}
\end{subequations}
The stress-energy tensor $T_{\mu \nu}$ is the sum of two contributions, one from the electromagnetic field and one from the scalar field:
\begin{equation}
T_{\mu \nu } = T_{\mu \nu }^F + T_{\mu \nu }^{\Phi},
\end{equation}
where
\begin{subequations}
\begin{align}
T_{\mu \nu }^{F} &= F_{\mu  \rho } F_{\nu }{}^{\rho } - \frac{1}{4} g_{\mu \nu } F_{\rho \sigma } F^{\rho \sigma }, \\
T_{\mu \nu }^{\Phi} &= D^\ast_{(\mu } \Phi^\ast D^{}_{\nu )} \Phi  -\frac{1}{2} g_{\mu \nu }\left[ g^{\rho \sigma } D^\ast_{(\rho } \Phi^\ast D^{}_{\sigma )} \Phi \right] .
\end{align}
\end{subequations}
In (\ref{eq:MW}), the current is given by
\begin{equation}
J^\mu = \frac{iq}{2} \left[\Phi^\ast D^\mu \Phi - \Phi (D^\mu \Phi)^\ast \right] .
\label{eq:current}
\end{equation}
The field equations (\ref{eq:field}) are invariant under a $U(1)$ gauge transformation
\begin{equation}
\Phi \rightarrow e^{i \chi} \Phi, \quad \quad A_\mu \rightarrow A_\mu + q^{-1} \chi_{,\mu } ,
\label{eq:gauge-transform}
\end{equation}
for any (real) scalar field $\chi $.

In this paper we are interested in static, spherically symmetric, soliton solutions of the field equations (\ref{eq:field}) and linear, spherically symmetric perturbations of these static solutions. We therefore consider a spherically symmetric, metric ansatz of the form
\begin{equation}
ds^{2} = -f h \, dt^2 + f^{-1} dr^2 + r^2 \left( d\theta ^{2} + \sin ^{2}\theta\, d\varphi ^{2} \right) ,
\label{eq:metric}
\end{equation}
where the metric functions $f(t,r)$ and $h(t,r)$ depend on time $t$ and the radial coordinate $r$ only.
The complex scalar field $\Phi (t,r)$ also depends only on $t$ and $r$.
By virtue of spherical symmetry, we may set the electromagnetic potential components $A_{\theta }$ and $A_{\varphi }$ to vanish identically.
Making an appropriate gauge transformation (\ref{eq:gauge-transform}), we can also set $A_{r}\equiv 0$.
The electromagnetic potential therefore takes the form
\begin{equation}
A_\mu = [A_{0}(t,r),0,0,0] .
\end{equation}
We define new variables
\begin{equation}
\gamma =  fh^{1/2}, \qquad
\Psi = r \Phi , \qquad
E = A_{0}',
\label{eq:newvars}
\end{equation}
in terms of which the field equations (\ref{eq:field}) take the form \cite{Dolan:2015dha}:
\begin{subequations}
\label{eq:dyn}
\begin{align}
\frac{f'}{f} &= -\frac{r}{2 \gamma^2} \left( \tau + f E^2 \right) + \frac{1}{fr} (1 - f),
\label{eq:dyn1} \\
\frac{h'}{h} &= \frac{r\tau}{\gamma^2},
\label{eq:dyn2} \\
-\frac{\dot{f}}{f} &= r \text{Re} \left( \dot{\Phi}^\ast \Phi' \right) + r q A_0 \text{Im} \left( \Phi'^{\ast} \Phi \right) ,
\label{eq:dyn3} \\
\left(  \frac{r^2 A'_{0}}{h^{1/2}} \right)' &= \frac {r^{2}}{\gamma } \left[ q^2 |\Phi|^2 A_0 - q \text{Im}\left(\dot{\Phi} \Phi^\ast\right) \right],
\label{eq:dyn4} \\
\partial_t \left( \frac{r A'_{0}}{h^{1/2}} \right) &=  - qr \text{Im}( \gamma\Phi' \Phi^\ast) ,
\label{eq:dyn5} \\
0 &=  -\ddot{\Psi} + \frac{\dot{\gamma}}{\gamma} \dot{\Psi} + \gamma\left(\gamma\Psi'\right)' - \frac{\gamma\gamma'}{r} \Psi + 2 i q A_0 \dot{\Psi}
\nonumber \\
& \qquad + i q \dot{A}_0 \Psi
 - i q \frac{\dot{\gamma}}{\gamma} A_0 \Psi + q^2 A_0^2 \Psi .
\label{eq:dyn6}
\end{align}
\end{subequations}
In (\ref{eq:dyn}), a dot $\dot {}$ denotes differentiation with respect to time $t$ and a prime ${}'$ differentiation with respect to the radial coordinate $r$.
We have in addition defined the quantity $\tau $ in (\ref{eq:dyn1}, \ref{eq:dyn2}) by
\begin{equation}
\tau = |\dot{\Phi}|^2 + |\gamma\Phi'|^2 + q^2 A_0^2 |\Phi|^2 + 2 q A_0 \text{Im}(\Phi \dot{\Phi}^\ast).
\end{equation}

For static field configurations, the variables $f$, $h$, $A_{0}$ and $\Phi $ all depend only on $r$, and the time derivatives in (\ref{eq:dyn}) all vanish.
We also assume that the scalar field $\Phi = \phi (r)$ is real for static equilibrium solutions.
In this case the field equations (\ref{eq:dyn}) reduce to
\begin{subequations}
\label{eq:static}
\begin{align}
h' &= r \left[ \left(\frac{q A_0 \phi}{f} \right)^2 +  h (\phi')^2 \right],
\label{eq:hprime} \\
 E^2 &= -\frac{2}{r} \left[ f' h + \frac{1}{2} f h' + \frac{h}{r}(f - 1) \right],
\label{eq:fprime}  \\
0 &= f A_0'' +  \left( \frac{2f}{r} - \frac{f h'}{2h} \right) A_0' - q^2 \phi^2 A_0,
\label{eq:Aprimeprime} \\
0 &= f \phi'' + \left(\frac{2f}{r} + f' + \frac{f h'}{2 h}\right) \phi' + \frac{(q A_0)^2}{fh} \phi .
\label{eq:phiprimeprime}
\end{align}
\end{subequations}

If the scalar field $\phi $ is set to vanish, $\phi \equiv 0$, then (\ref{eq:Aprimeprime}) implies that $A_{0}\equiv 0$ if the electromagnetic potential $A_{0}$ is finite at the origin. Therefore the only trivial solution of the field equations (\ref{eq:static}) representing a soliton has vanishing scalar and electromagnetic field, and the metric is that of Minkowski spacetime. This is in contrast to the black hole case, where the charged Reissner-Nordstr\"om black hole is a solution of the field equations (\ref{eq:static}) with vanishing scalar field.

\section{Gravitating charged-scalar solitons in a cavity}
\label{sec:solitons}

In this section we consider static, spherically symmetric, soliton solutions of the field equations (\ref{eq:static}).
We require that all the field variables and all physical quantities (electromagnetic field strength, curvature, etc) are regular at the origin $r=0$.
With this condition, the field variables have the following expansions for small $r$:
\begin{align}
f &= 1-  \left(\frac{q^2{\phi_0}^2{a_0}^2}{6h_0}\right)r^2 + O(r^3),
\nonumber \\
h &= h_0 + \left(\frac{q^2 {\phi_0}^2{a_0}^2}{2}\right)r^2 + O(r^3),
\nonumber \\
A_0 &= a_0 + \left(\frac{a_0q^2{\phi_0}^2}{6}\right)r^2 + O(r^3),
\nonumber \\
\phi &= \phi_0 - \left(\frac{\phi_0q^2{a_0}^2}{6h_0}\right)r^2 + O(r^3).
\label{eq:origin}
\end{align}
Here, $\phi _{0}$, $a_{0}$ and $h_{0}$ are arbitrary constants, with $h_{0}>0$ so that the metric (\ref{eq:metric}) has the correct signature.

It is straightforward to show, using an adaptation of the argument in \cite{Bekenstein:1971hc}, that there are no nontrivial asymptotically flat soliton solutions of the field equations (\ref{eq:static}) -- see Appendix \ref{sec:appendix}.
In analogy with the black hole solutions found in \cite{Dolan:2015dha}, we therefore consider soliton solutions in a cavity, with a reflecting mirror at
$r=r_{m}$. At the mirror the scalar field must vanish, so
\begin{equation}
\phi (r_{m}) = 0.
\label{eq:mirror}
\end{equation}

The static field equations (\ref{eq:static}) possess two scaling symmetries.  Firstly, there is a length scaling symmetry. Define new variables $R$, $Q$ as follows:
\begin{equation}
r = LR, \qquad q = L^{-1}Q,
\label{eq:newvars1}
\end{equation}
where $L$ is an arbitrary constant length scale, and $f$, $h$, $\phi $ and $A_{0}$ are unchanged.
Substituting (\ref{eq:newvars1}) into the static field equations (\ref{eq:static}) yields
\begin{align}
\frac {dh}{dR} &= R \left[ \left(\frac{Q A_0 \phi}{f} \right)^2 + h\left( \frac {d\phi }{dR} \right)^2 \right],
\nonumber \\
\left( \frac {dA_{0}}{dR} \right) ^{2}  &= -\frac{2}{R} \left[ \frac {df}{dR} h + \frac{1}{2} f \frac {dh}{dR} + \frac{h}{R}(f - 1) \right],
\nonumber  \\
0 &= f \frac {d^{2}A_0}{dR^{2}} +  \left( \frac{2f}{R} - \frac{f}{2h}\frac {dh}{dR} \right) \frac {dA_0}{dR}
\nonumber \\ & \qquad
- Q^2 \phi^2 A_0,
\nonumber \\
0 &= f \frac {d^{2}\phi}{dR^{2}} + \left(\frac{2f}{R} + \frac {df}{dR} + \frac{f}{2h}\frac {dh}{dR}\right) \frac {d\phi }{dR}
\nonumber \\ & \qquad + \frac{ (Q A_0)^2}{fh} \phi ,
\end{align}
which are identical to the original equations (\ref{eq:static}).
Secondly, we can rescale the time coordinate. In this case we define new variables $H$ and ${\mathfrak {A}}_{0}$ as follows:
\begin{equation}
h=T^{-2}H, \qquad A_{0} = T^{-1} {\mathfrak {A}}_{0},
\label{eq:newvars2}
\end{equation}
where $T$ is an arbitrary constant time scale and $f$, $\phi $ and $q$ are unchanged.
Substituting (\ref{eq:newvars2}) into the static field equations (\ref{eq:static}) gives the equations (with $'$ denoting differentiation with respect to $r$) \begin{align}
H' &= r \left[ \left(\frac{q {\mathfrak {A}}_0 \phi}{f} \right)^2 +  H (\phi')^2 \right],
\nonumber \\
\left( {\mathfrak {A}}_{0}' \right) ^{2} &= -\frac{2}{r} \left[ f' H + \frac{1}{2} f H' + \frac{H}{r}(f - 1) \right],
 \nonumber \\
0 &= f {\mathfrak {A}}_0'' +  \left( \frac{2f}{r} - \frac{f H'}{2H} \right) {\mathfrak {A}}_0' - q^2 \phi^2 {\mathfrak {A}}_0,
\nonumber \\
0 &= f \phi'' + \left(\frac{2f}{r} + f' + \frac{f H'}{2 H}\right) \phi' + \frac{(q {\mathfrak {A}}_0)^2}{fH} \phi ,
\end{align}
which are again identical to the original static field equations (\ref{eq:static}).

\begin{figure}
\includegraphics[width=8.5cm]{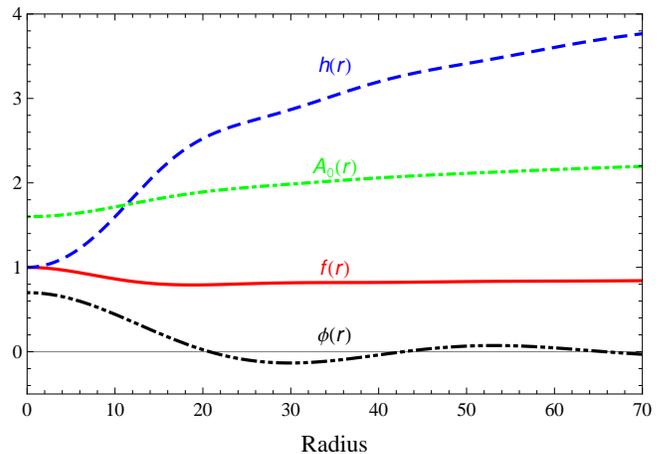}
\caption{A typical soliton solution with scalar charge $q=0.1$, $a_{0}=1.6$ and $\phi _{0}=0.7$.  We plot the metric functions $f(r)$, $h(r)$ and matter field functions $A_{0}(r)$, $\phi (r)$.}
\label{fig:one}
\end{figure}

We use the time-coordinate rescaling (\ref{eq:newvars2}) to set $h(0)=h_{0}=1$ without loss of generality.
The expansions (\ref{eq:origin}) are then determined by the parameters $q$, $a_{0}$ and $\phi _{0}$.
We use the length rescaling (\ref{eq:newvars1}) to set $q=0.1$, leaving the free parameters $a_{0}$ and $\phi _{0}$. We choose $q=0.1$ to facilitate comparison with the black hole solutions presented in \cite{Dolan:2015dha}, where the length rescaling (\ref{eq:newvars1}) was used to fix the radius of the black hole event horizon to be unity, so that $q$ was a free parameter in that case.

\begin{figure}
\includegraphics[width=8.5cm]{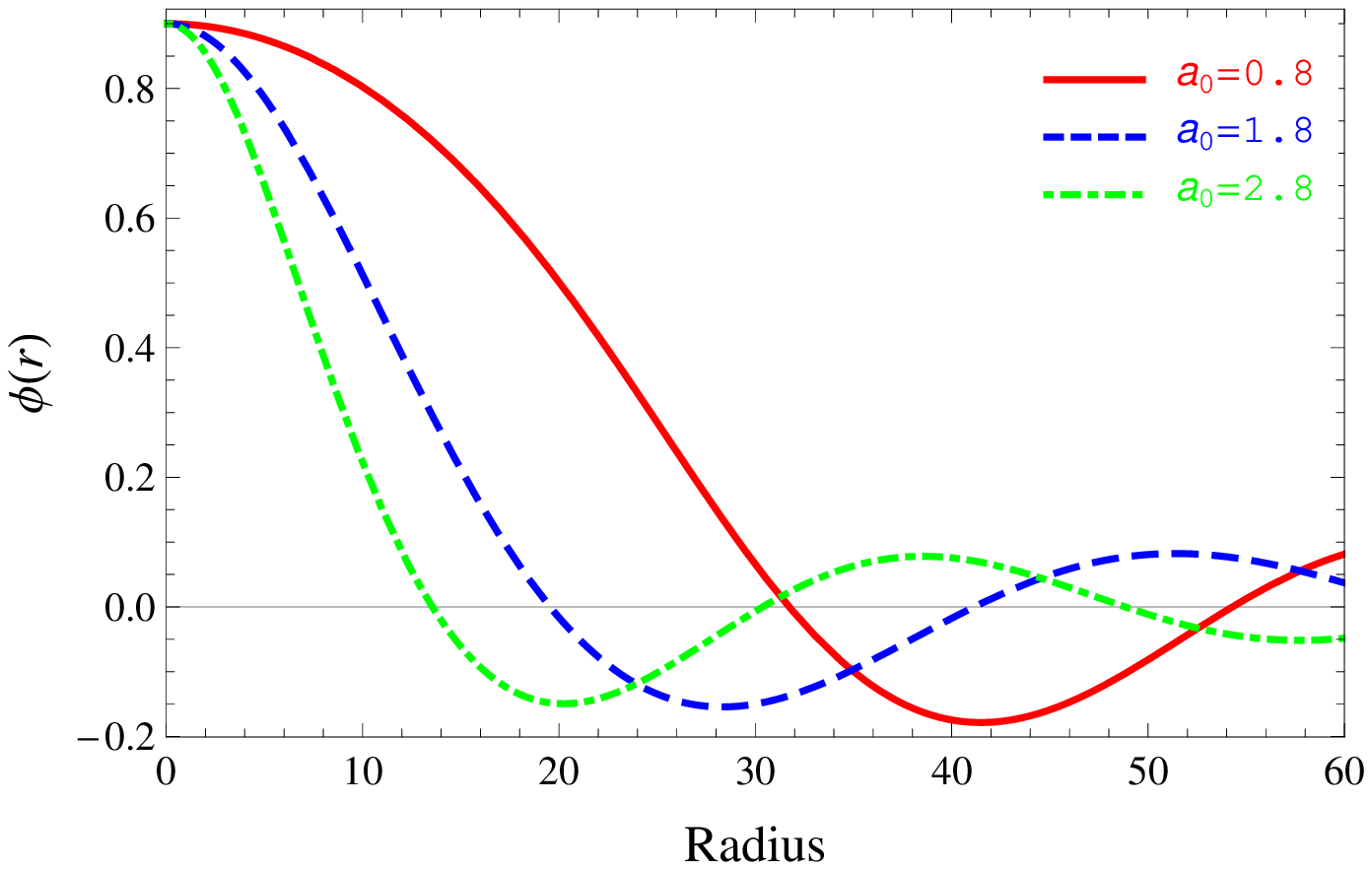}
\includegraphics[width=8.5cm]{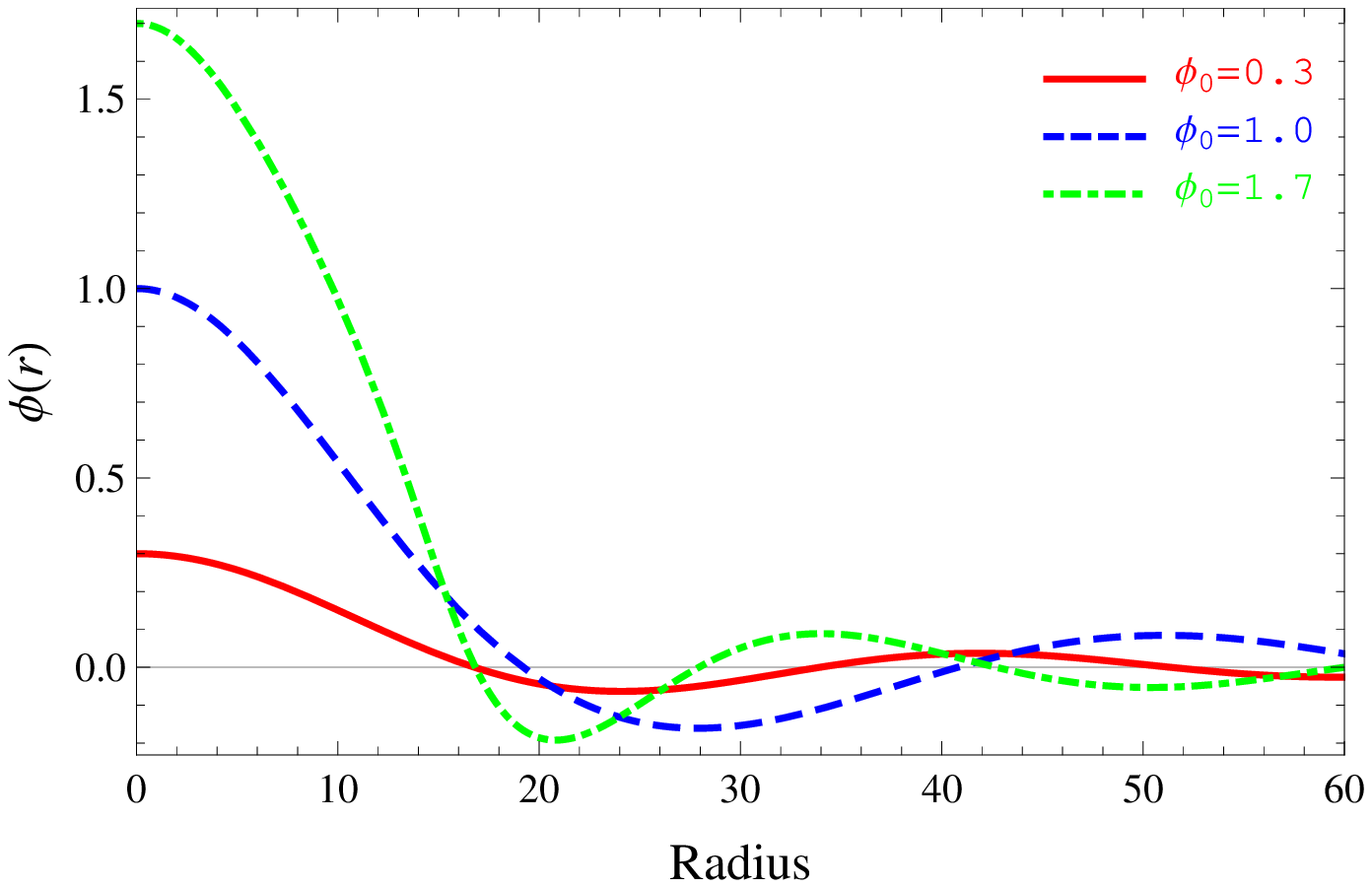}
\caption{Scalar field profiles for static soliton solutions with fixed scalar charge $q=0.1$. Top: Fixed $\phi _{0}=0.9$ and three distinct values of $a_{0}$. Bottom:
Fixed $a_{0}=1.9$ and three distinct values of $\phi _{0}$.}
\label{fig:two}
\end{figure}

\begin{figure}
\includegraphics[width=8.5cm]{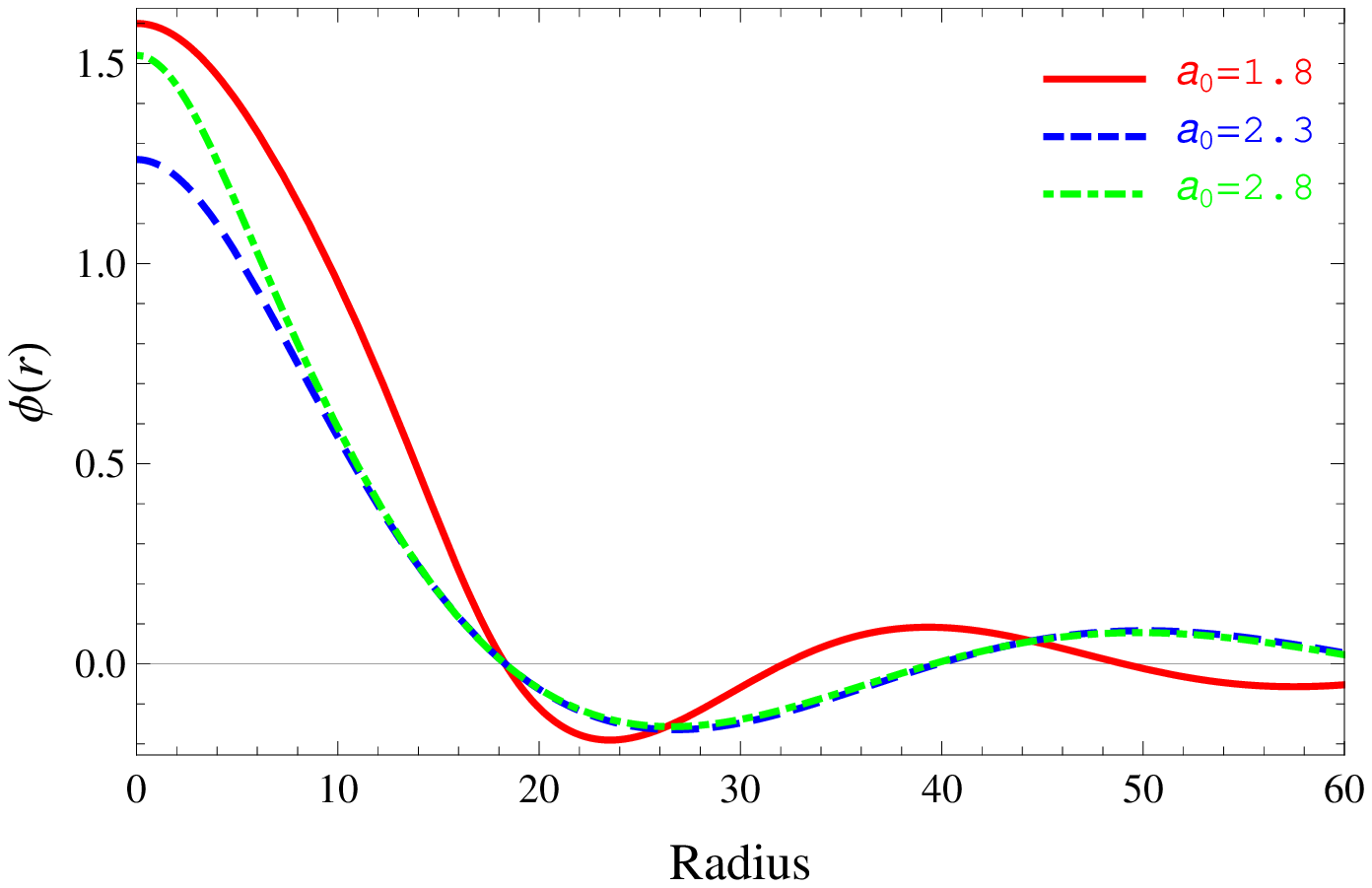}
\includegraphics[width=8.5cm]{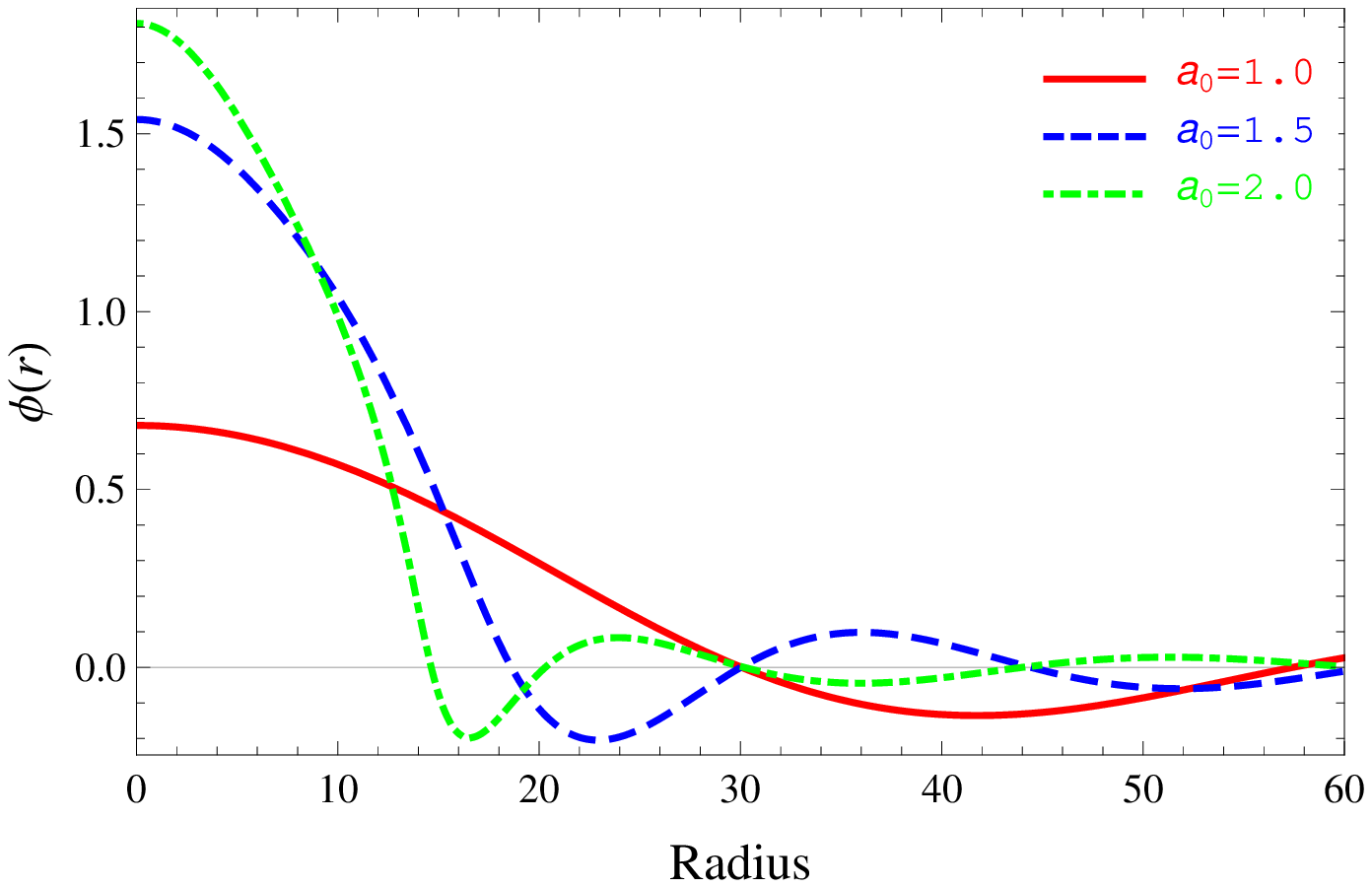}
\caption{Scalar field profiles for soliton solutions with scalar charge $q=0.1$. Top: Three scalar field profiles which share the same location of their first zero at $r_{m}\approx 18$. Bottom:  Three scalar field profiles with a common zero: the first (red, solid curve), second (blue, dashed curve) and third (green, dotted curve) zeros are at $r_{m} \approx 30$.}
\label{fig:three}
\end{figure}

The static field equations (\ref{eq:static}) are integrated numerically to find soliton solutions.  We start the numerical integration at $r=\epsilon $, where $\epsilon $ is typically $10^{-12}$, using the expansions (\ref{eq:origin}) as initial conditions.
In Fig.~\ref{fig:one} we plot the four field variables $f(r)$, $h(r)$, $A_{0}(r)$ and $\phi (r)$ for a typical soliton solution with $q=0.1$, $a_{0}=1.6$ and $\phi _{0}=0.7$.
As for the black hole solutions in \cite{Dolan:2015dha}, the scalar field $\phi $ oscillates about zero;  the mirror can be placed at any zero of the scalar field. In this paper, we  consider the case where the mirror is located at the first zero of $\phi $.
This is because the black hole solutions studied in \cite{Dolan:2015dha} were stable under linear, spherically symmetric perturbations when the mirror was at the first zero of the scalar field, but unstable when the mirror was at the second zero of the scalar field.

Various scalar field profiles for soliton solutions are shown in Fig.~\ref{fig:two}, where the oscillatory behaviour of the scalar field can be clearly seen.  In the upper plot, we fix $\phi _{0}=0.9$ and show the profiles for three values of $a_{0}$; in the lower plot we fix $a_{0}=1.9$ and show the profiles for three values of $\phi _{0}$.
We see that for this fixed value of $\phi _{0}$, the radius of the first node of the scalar field decreases as $a_{0}$ increases, while the behaviour of the location of the first zero of $\phi$ for fixed $a_{0}$ and varying $\phi _{0}$ is more complicated.

\begin{figure}
\includegraphics[width=8.5cm]{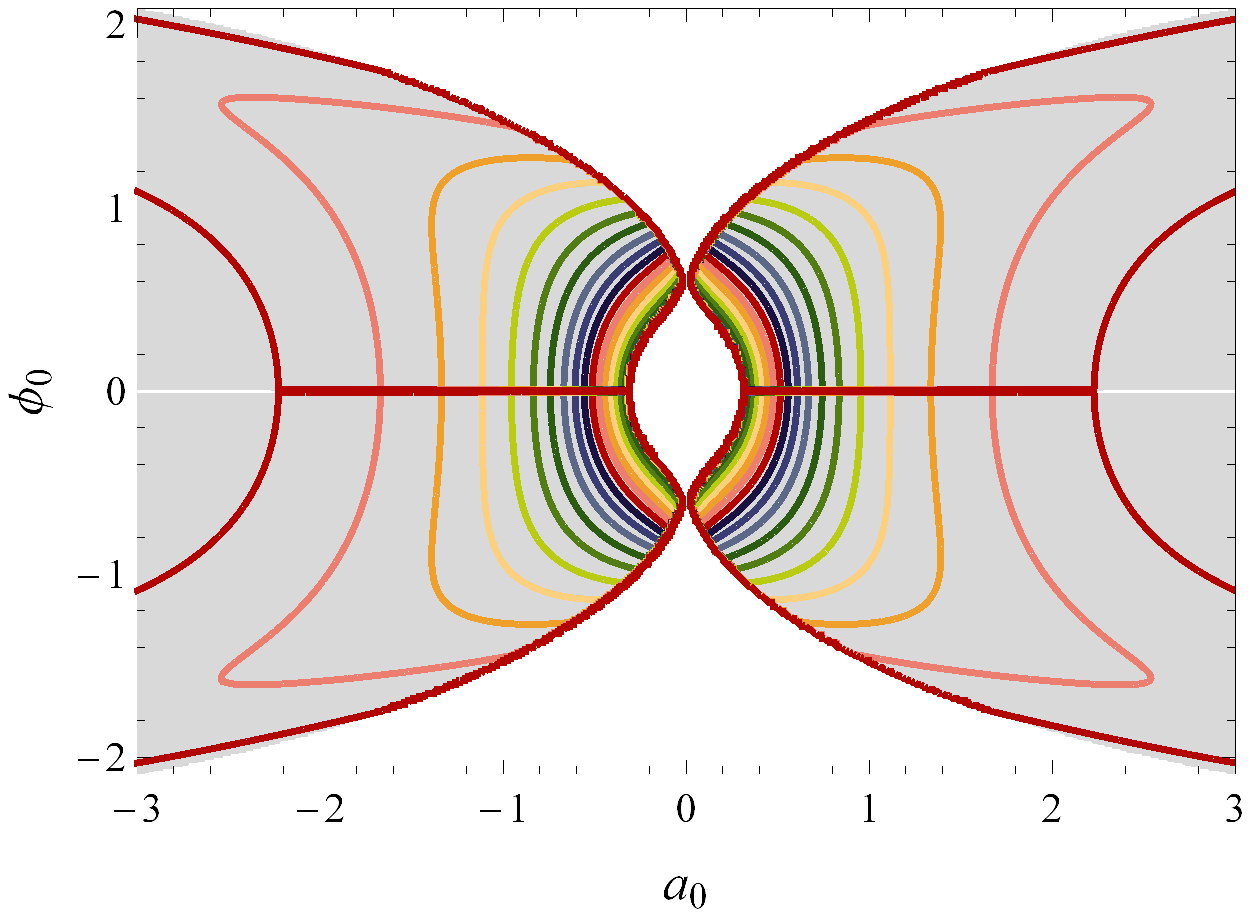}
\caption{Portion of the phase space of charged scalar soliton solutions in a cavity with scalar charge $q=0.1$. The solutions are described by two parameters: $a_{0}$ (horizontal axis) and $\phi _{0}$ (vertical axis).  Solutions exist in the shaded regions. The mirror radius $r_{m}$ is assumed to be at the first zero of the scalar field and the shaded region denotes solutions with $r_{m}\le 100$. The lines are contours of constant $r_{m}$. There are no solutions on the axes $a_{0}=0$ or $\phi _{0}=0$. Solutions also exist in the central region of the plot, towards $a_{0}\rightarrow 0$ and $\phi _{0}\rightarrow 0$, where the mirror radius $r_{m}>100$.
The values of the mirror radius $r_{m}$ are given for selected contours.}
\label{fig:four}
\end{figure}

It is possible to have two (or more) solitons with the same mirror radius $r_{m}$, as shown in Fig.~\ref{fig:three} (this behaviour was also found for the black hole solutions \cite{Dolan:2015dha}).
The top plot in Fig.~\ref{fig:three} shows three scalar field profiles which have the same first zero at $r_{m}\approx 18$, with different values of $a_{0}$ and $\phi _{0}$.  The lower plot in Fig.~\ref{fig:three} shows a further three scalar field profiles, again with different values of $a_{0}$ and $\phi _{0}$, whose first, second and third zeros respectively lie at $r_{m} \approx 30$.

A portion of the phase space of solutions is shown in Fig.~\ref{fig:four}.  We fix the scalar charge $q=0.1$, although the phase space is independent of $q$ due to the scaling symmetry (\ref{eq:newvars1}). The overall structure of the phase space shares many features with that for black hole solutions, shown in Ref.~\cite{Dolan:2015dha}, but with some notable differences as well.
With fixed $q$, the phase space depends on the two parameters $a_{0}$ and $\phi _{0}$. There are no nontrivial solutions when either $a_{0}=0$ or $\phi _{0}=0$. We consider values of $a_{0}$ between -3 and +3.
The shaded region in Fig.~\ref{fig:four} shows where soliton solutions exist with $r_m\le 100$, when the mirror is placed at the first zero of the scalar field.
There are also solutions in the central region towards $a_{0}\rightarrow 0$ and $\phi _{0}\rightarrow 0$, with $r_{m}>100$.
The mirror radius $r_{m}$ generally decreases as we move away from the origin.

For black hole solutions, the requirement of a regular event horizon at $r=r_{h}$ restricts the phase space (in particular, the value of $A_{0}'$ on the event horizon has an upper bound of ${\sqrt {2}}/r_{h}$ for $h(r_{h})=1$, see Fig.~4 in \cite{Dolan:2015dha}), but for soliton solutions we have no {\it {a priori}} restrictions on the values of either $a_{0}$ or $\phi _{0}$.
For each fixed value of $a_{0}$, we find nontrivial soliton solutions when $\phi _{0}$ lies in some bounded interval; outside this interval the metric function $f$ either has a zero or the solution becomes singular before the scalar field $\phi $ has a zero.
However, we find no upper limit on the value of $a_{0}$ for which there exist nontrivial solutions - only a portion of the phase space is shown in Fig.~\ref{fig:four}.  For large $a_{0}$ we find that the mirror radius is extremely small. For example, with $a_{0}=10^{5}$ and $\phi _{0}=1$,
there exists a soliton solution with $r_{m}\approx 4\times 10^{-4}$.

\begin{figure}
\includegraphics[width=8.5cm]{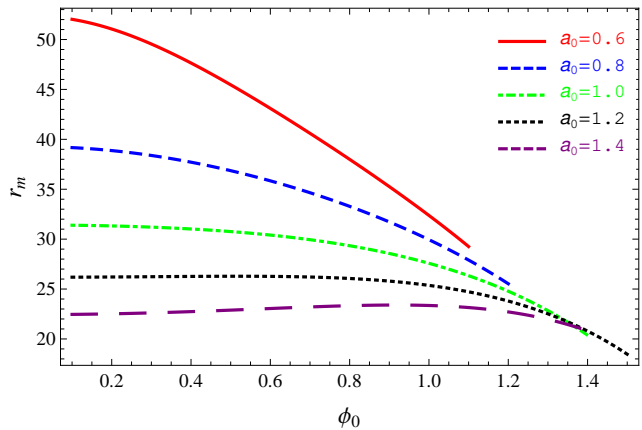}
\caption{Location of the mirror $r_{m}$ at the first zero of the scalar field, with scalar charge $q=0.1$, as a function of $\phi _{0}$ for various fixed values of $a_{0}$.}
\label{fig:five}
\end{figure}

\begin{figure*}
\includegraphics[width=8.5cm]{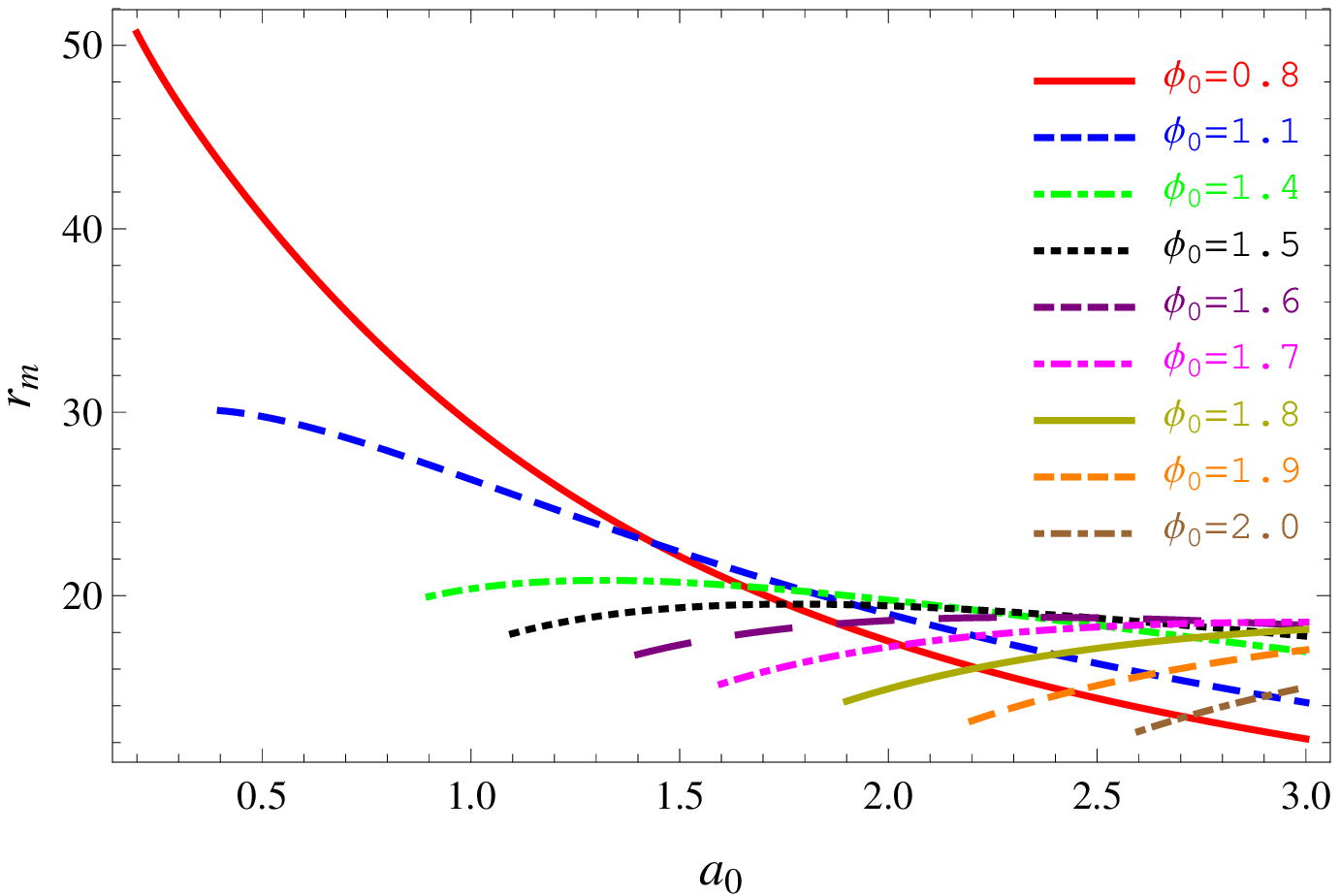}
\includegraphics[width=8.5cm]{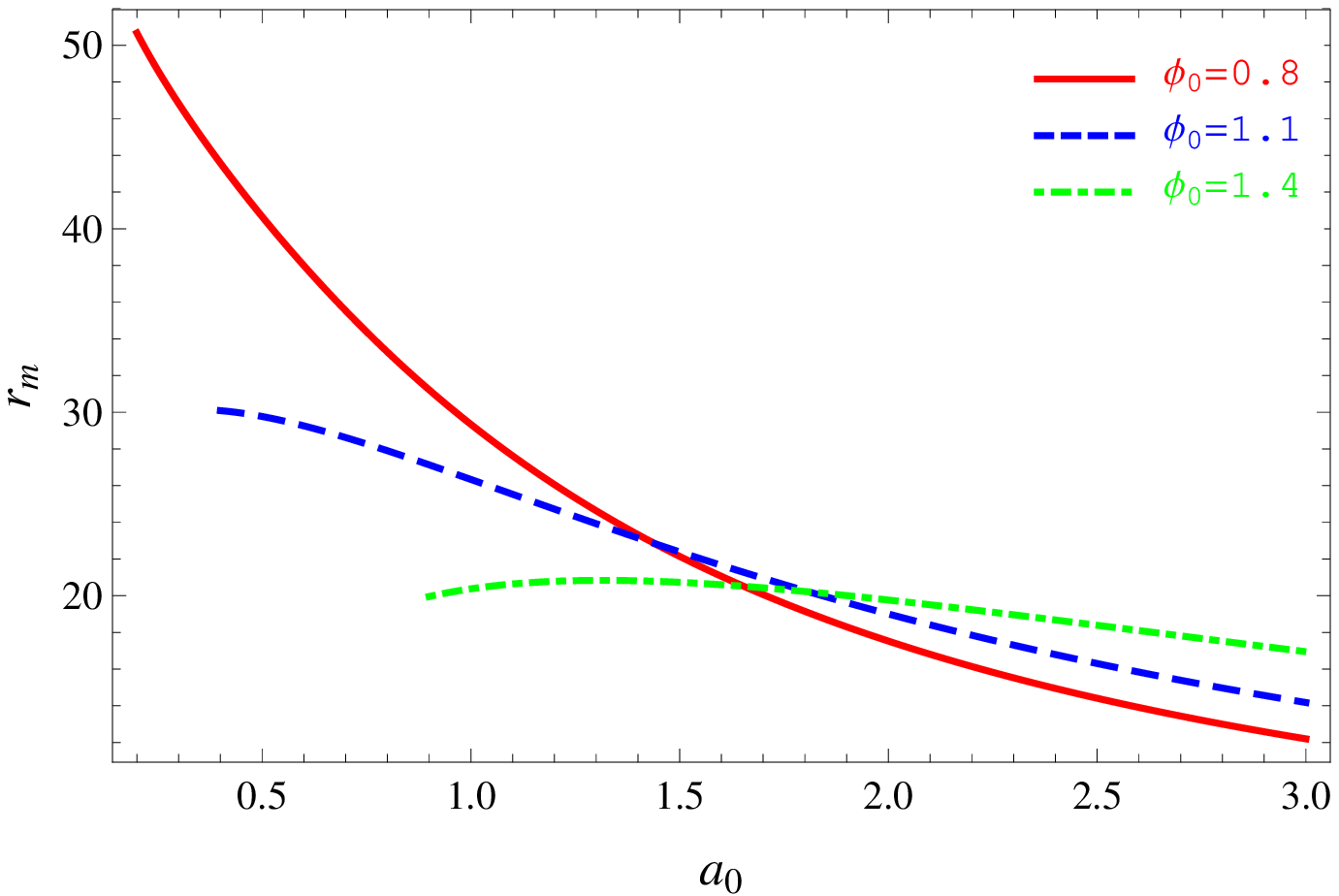}
\includegraphics[width=8.5cm]{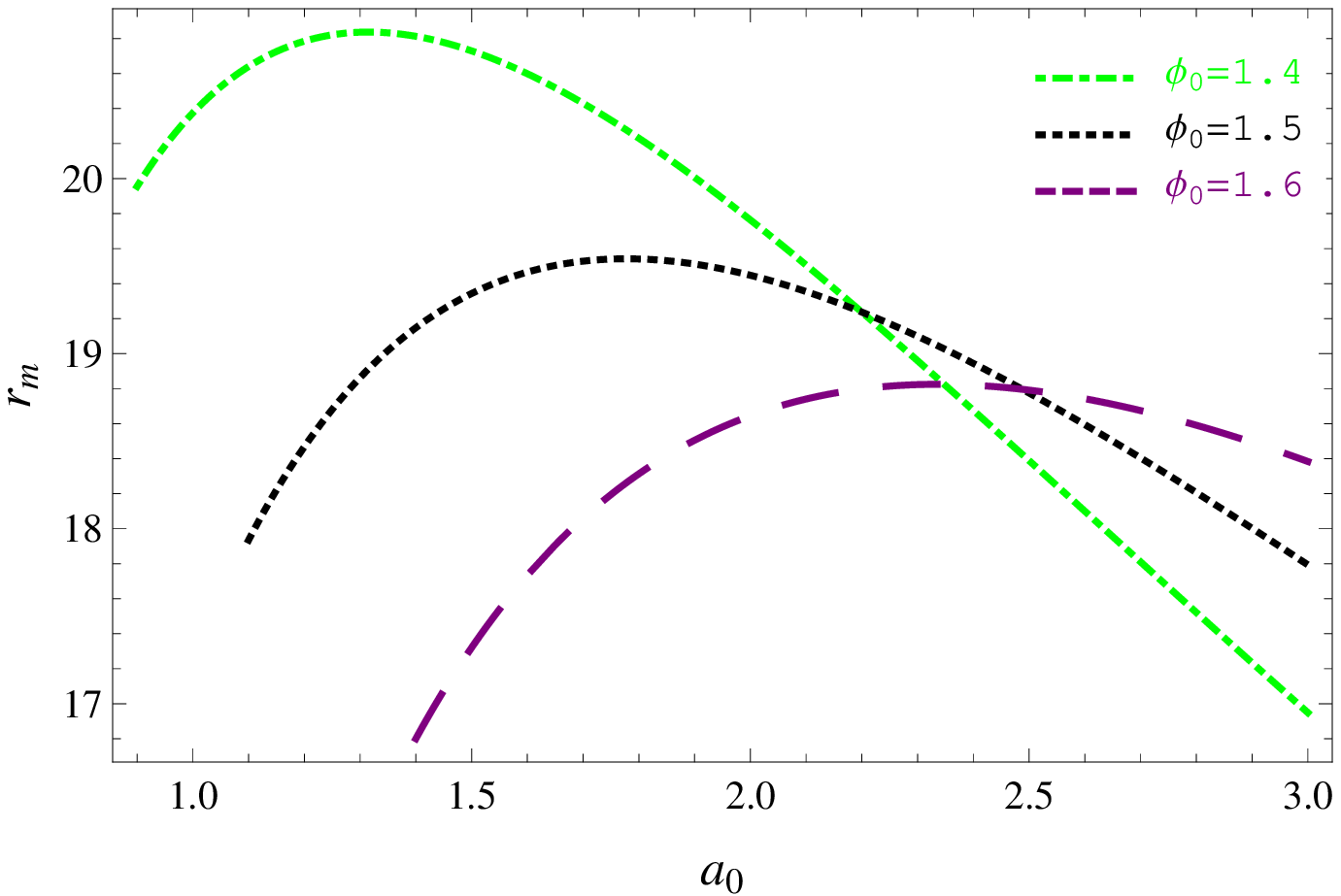}
\includegraphics[width=8.5cm]{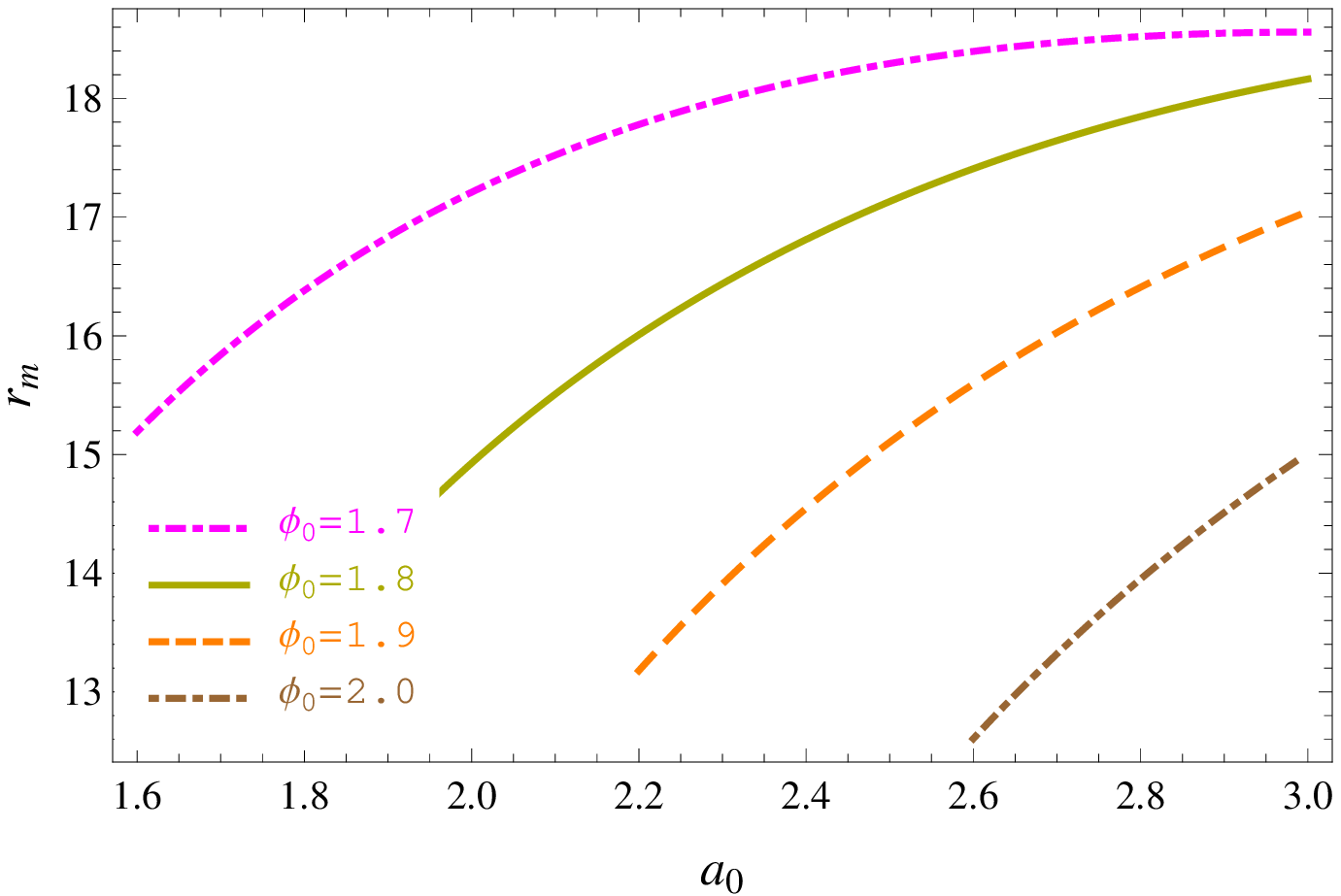}
\caption{Location of the mirror $r_{m}$ at the first zero of the scalar field, with scalar charge $q=0.1$, as a function of $a_{0}$ for various fixed values of $\phi _{0}$. Top left: $r_{m}$ for $\phi _{0}\in [0.8, 2.0]$.  To make the behaviour more visible, the data in the top-left plot is repeated in the remaining plots, just for a few values of $\phi _{0}$.}
\label{fig:six}
\end{figure*}

In Figs.~\ref{fig:five} and \ref{fig:six} we explore how the mirror radius $r_{m}$ (when the mirror is at the first zero of the scalar field) depends on the parameters $a_{0}$ and $\phi _{0}$.
We plot in Fig.~\ref{fig:five} the radius $r_{m}$ as a function of $a_{0}$ for various fixed values of $\phi _{0}$.
For smaller fixed values of $a_{0}$, the location of the mirror $r_{m}$ decreases as $\phi _{0}$ increases, but for larger fixed $a_{0}$, the mirror radius increases slightly as $\phi _{0}$ increases before decreasing again.
For fixed $\phi _{0}$ and various values of $a_{0}$, the radius $r_{m}$ is shown in Fig.~\ref{fig:six}.
For smaller fixed values of $\phi _{0}$ (top right plot in Fig.~\ref{fig:six}), we see that $r_{m}$ decreases as $a_{0}$ increases; for large fixed values of $\phi _{0}$ (bottom right plot in Fig.~\ref{fig:six}) $r_{m}$ increases as $a_{0}$ increases (at least for the values of $a_{0}$ shown); while for intermediate values of $\phi _{0}$ (bottom left plot in Fig.~\ref{fig:six}), it can be seen that $r_{m}$ first increases to a maximum value then decreases as $a_{0}$ increases.

\begin{figure}
\includegraphics[width=8.5cm]{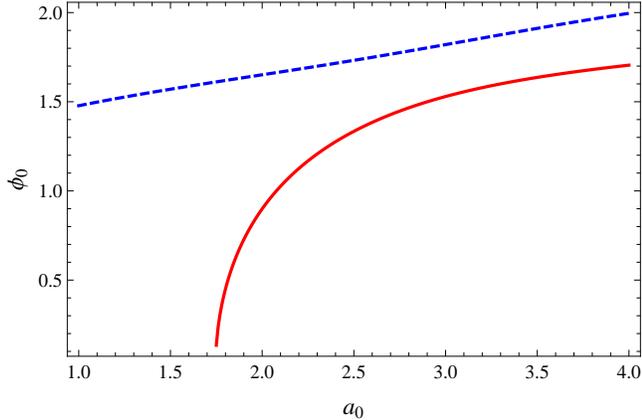}
\caption{Portions of the contour in the phase space of charged-scalar solitons in a cavity with $r_{m}=18$ and scalar charge $q=0.1$. The mirror is at the first zero of the scalar field.  There are two branches of solutions with this mirror radius; for fixed $a_{0}$ one branch has a larger value of $\phi _{0}$ than the other.}
\label{fig:seven}
\end{figure}

The contours of constant $r_{m}$ in the $(a_{0},\phi _{0})$-plane shown in Fig.~\ref{fig:four} also exhibit complicated behaviour.
For larger values of $r_{m} \gtrsim 20$, the contours are a single curve in each quadrant of the plane, such that there is one value of $\phi _{0}>0$ for each value of $a_{0}>0$ on the contour.
Similar behaviour is seen in the contours of constant $r_{m}$ for the black holes studied in \cite{Dolan:2015dha}.
However, for smaller values of $r_{m} \lesssim 20$, the contours of constant $r_{m}$ have two parts in each quadrant: the first part starts at some value of $\phi _{0}>0$ at the upper boundary of the phase space and has $\phi _{0}$ slowly increasing as $a_{0}$ increases; the second part begins at a small value of $\phi _{0}>0$ close to the horizontal axis and $\phi _{0}$ increases rapidly as $a_{0}$ increases along the contour. For the $r_{m}=19$ contour plotted in Fig.~\ref{fig:four}, the two branches meet at a larger value of $a_{0}$.
This behaviour is not seen in the phase spaces of black hole solutions \cite{Dolan:2015dha}, as these have a restricted range of values of $A_{0}'$ on the horizon.
To illustrate these two branches, in Fig.~\ref{fig:seven} we show a portion of the contour for charged soliton solutions with $r_{m}=18$.
It can be clearly seen that in this region of parameter space there are two branches of solitons with the same mirror radius (with the mirror at the first zero of the scalar field). For fixed $a_{0}$, one of these branches has a larger value of $\phi _{0}$ than the other.  We expect that if we continued the curves to larger values of $a_{0}$, ultimately the two branches might join together.

\begin{figure*}
\includegraphics[width=8.5cm]{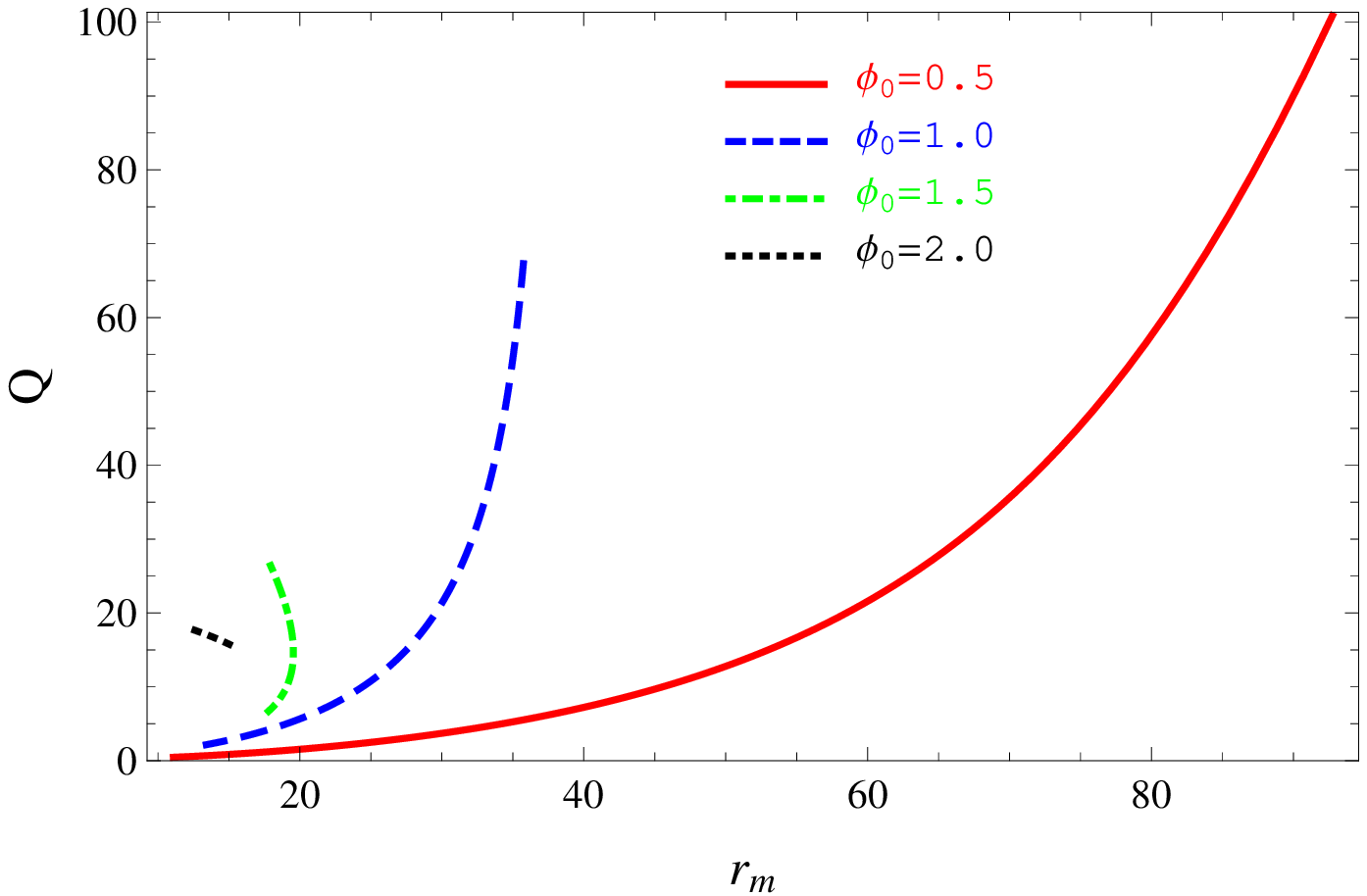}
\includegraphics[width=8.5cm]{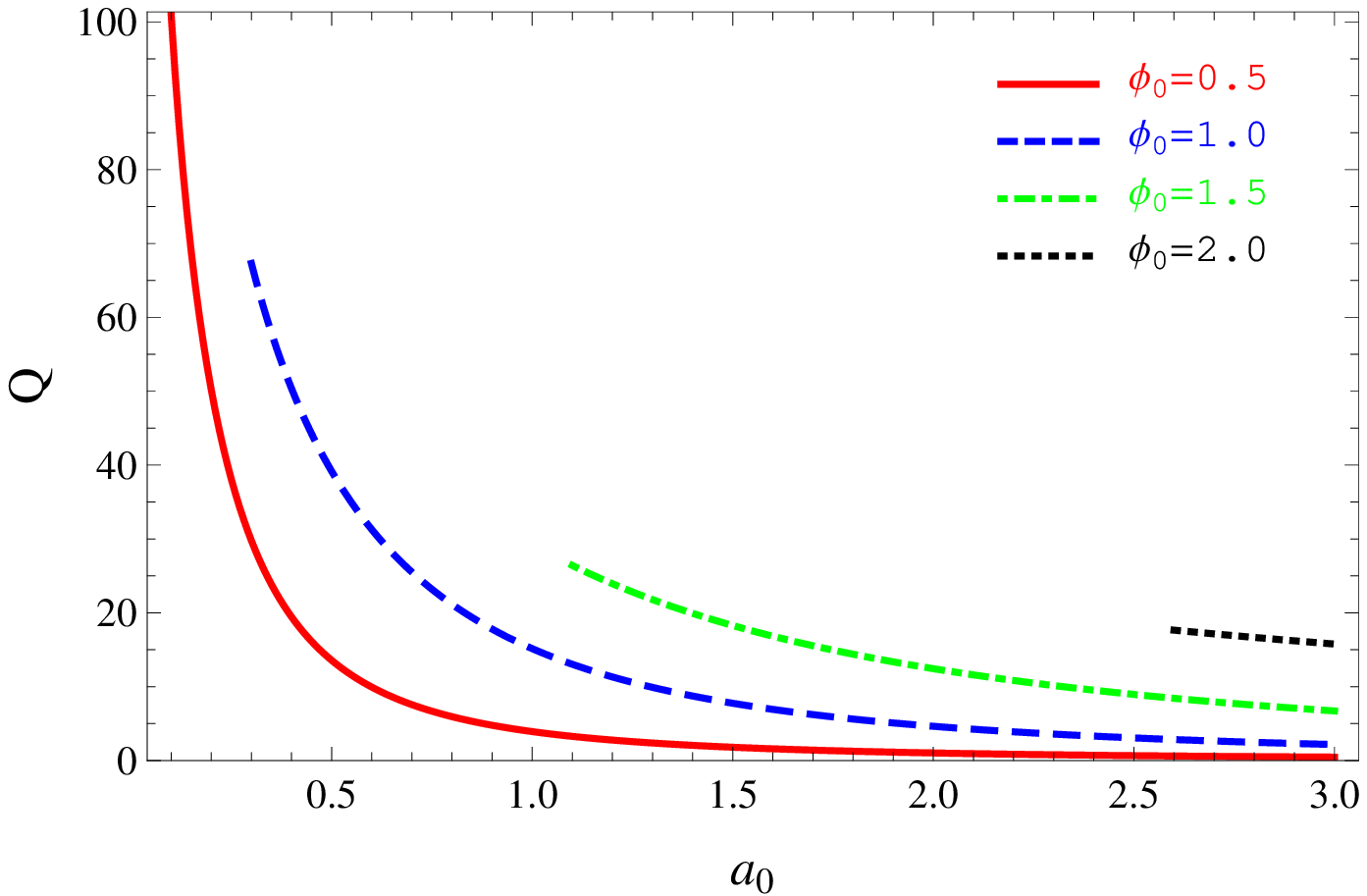}
\includegraphics[width=8.5cm]{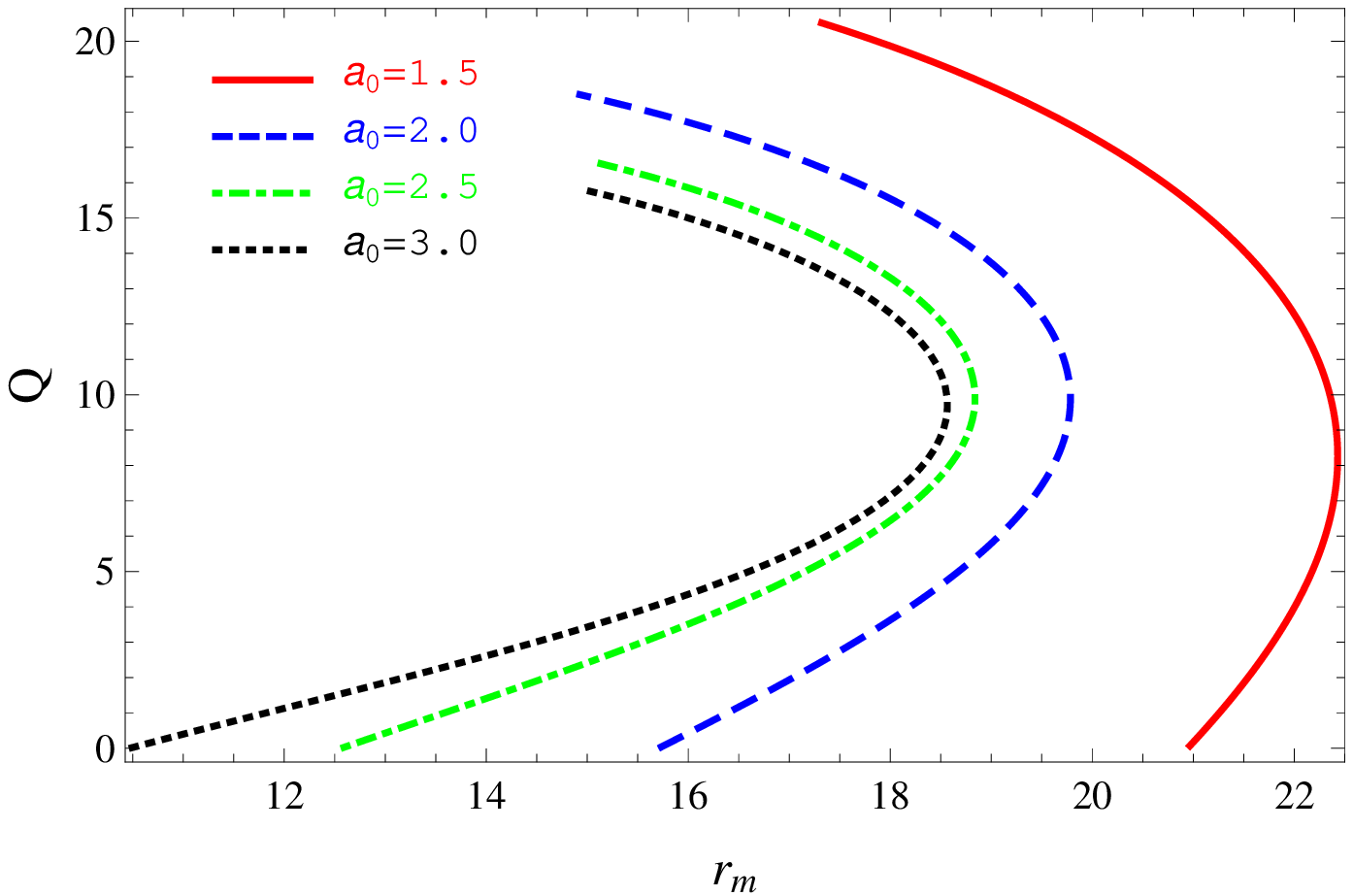}
\includegraphics[width=8.5cm]{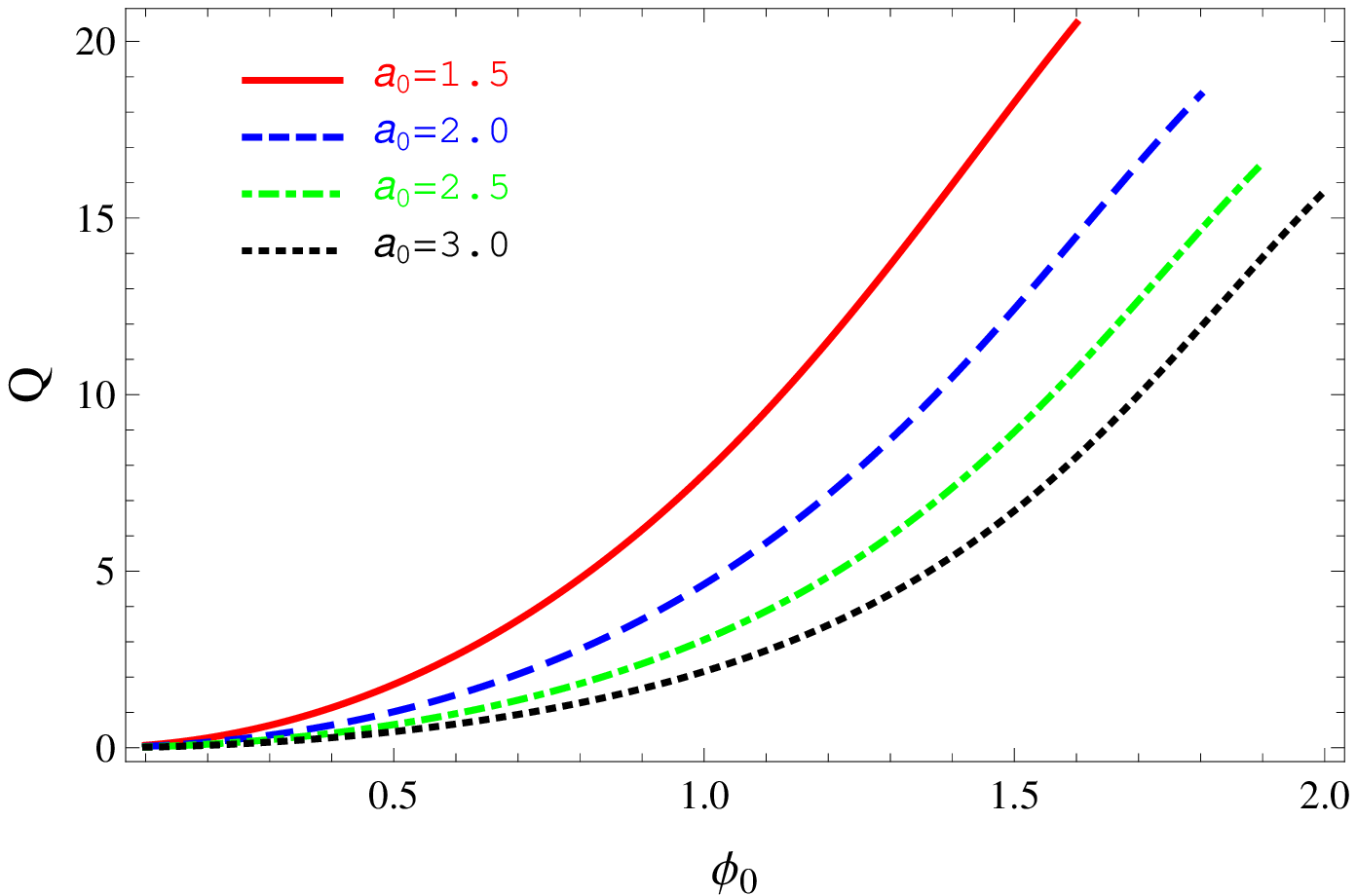}
\caption{Soliton electric charge $Q$ (\ref{eq:charge}) as a function of the parameters $a_{0}$ and $\phi _{0}$ for scalar charge $q=0.1$. The mirror is at the first zero of the scalar field.  Top row:  various fixed values of $\phi _{0}$ and $a_{0}\in [0.1, 3.0]$.  Bottom row: various fixed values of $a_{0}$ and $\phi _{0}\in [0.1,2.0]$.
Left-hand plots: $Q$ as a function of the mirror radius $r_{m}$.  Right-hand plots: the same data as the left-hand plots, but with $Q$ as a function of either $a_{0}$ or $\phi _{0}$, as applicable.}
\label{fig:eight}
\end{figure*}

\begin{figure*}
\includegraphics[width=8.5cm]{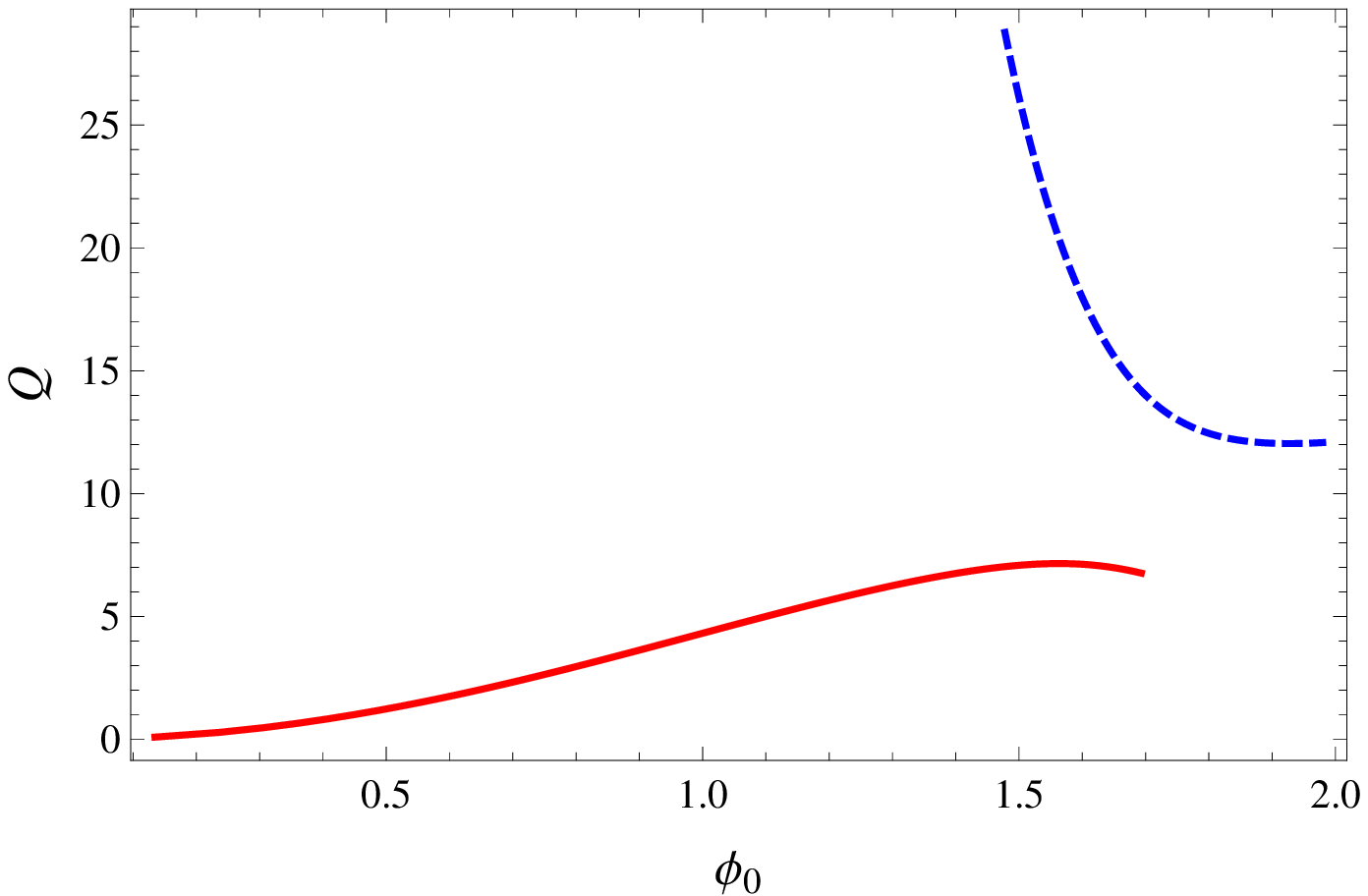}
\includegraphics[width=8.5cm]{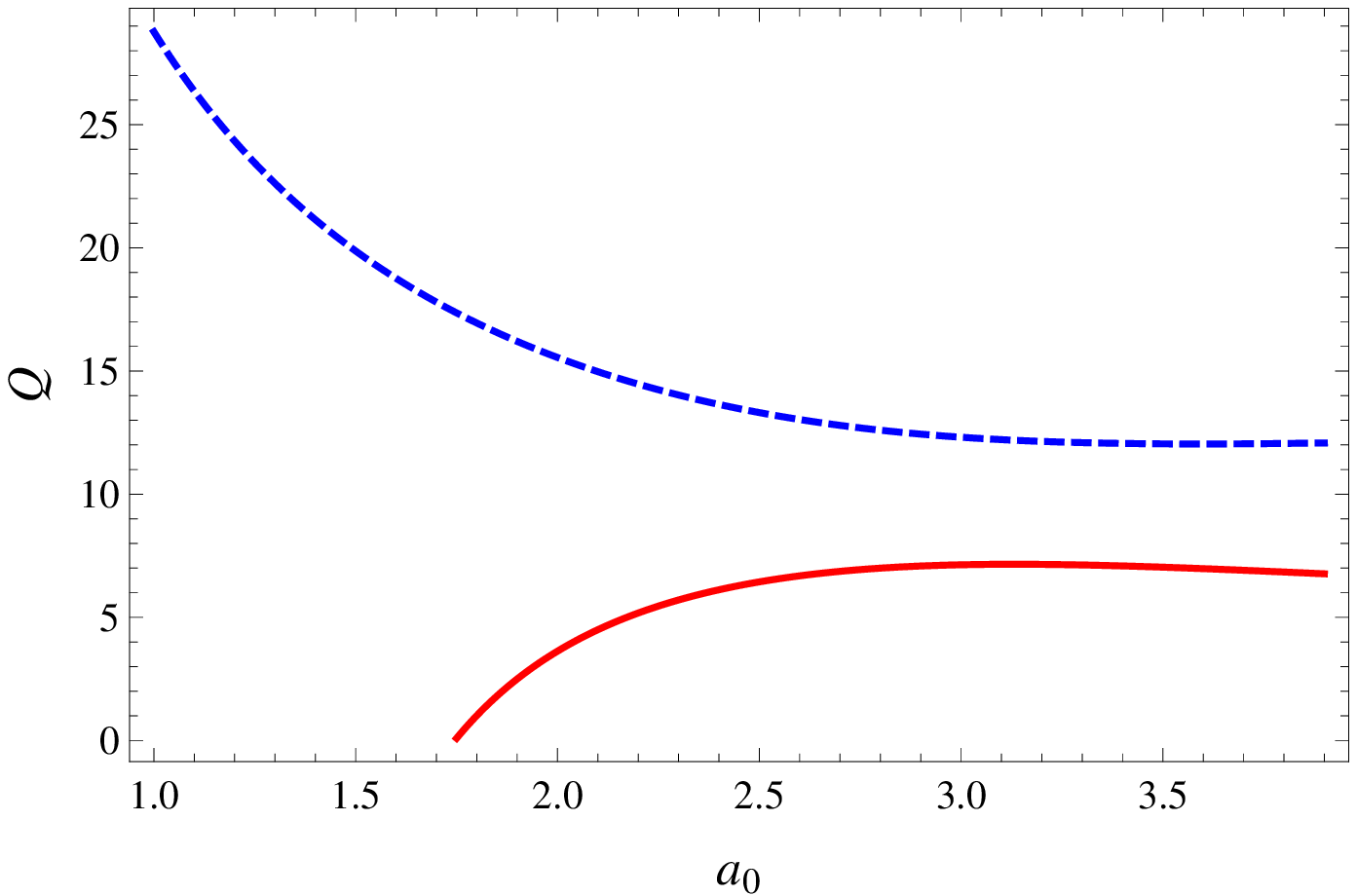}
\caption{Soliton electric charge $Q$ (\ref{eq:charge}) as a function of the parameters $a_{0}$ and $\phi _{0}$ for fixed mirror radius with scalar charge $q=0.1$. The mirror is at the first zero of the scalar field and is fixed to be at radius $r_{m}=18$. The blue (dashed) and red (solid) curves correspond to the two parts of the $r_{m}=18$ contour shown in Fig.~\ref{fig:seven}. The two plots show the same data, but plotted as a function of the different parameters $\phi _{0}$ and $a_{0}$.}
\label{fig:nine}
\end{figure*}

The electromagnetic current $J^{\mu }$ (\ref{eq:current}) is conserved, $\nabla _{\mu } J^{\mu }=0$.
Therefore, for each static equilibrium solution, we may define its electric charge by
\begin{equation}
Q = - \frac {1}{4\pi } \int _{\Sigma} d^{3}x {\sqrt {g^{(3)}}} n_{\mu }J^{\mu },
\end{equation}
where the integral is performed over a $t={\text{constant}}$ hypersurface $\Sigma $ with unit normal $n^{\mu }$, on which the induced metric has determinant $g^{(3)}$.
Performing the integral, we obtain
\begin{equation}
Q = - \frac {r_{m}^{2} A_{0}'(r_{m})}{{\sqrt {h(r_{m})}}}.
\label{eq:charge}
\end{equation}
This expression reduces to the usual definition of electric charge for the Reissner-Nordstr\"om black hole on taking $r_{m}\rightarrow \infty $.
In Figs.~\ref{fig:eight} and \ref{fig:nine} we explore how the charge $Q$ of the solitons depends on the parameters $\phi _{0}$ and $a_{0}$ and the mirror radius $r_{m}$.

Firstly, in Fig.~\ref{fig:eight} we consider the charge $Q$ for fixed values of $\phi _{0}$ and varying $a_{0}$; and for fixed values of $a_{0}$ and varying $\phi _{0}$. We plot the data for $Q$ as a function of either $\phi _{0}$ or $a_{0}$ (as applicable) and the same data also as a function of mirror radius $r_{m}$.
For fixed $\phi _{0}$, we see from the top-right plot that $Q$ decreases as $a_{0}$ increases.  The curve for $\phi _{0}=2.0$ is very short because we only consider values of $a_{0}$ up to $3.0$.  It will extend if we include larger values of $a_{0}$. With $a_{0}$ fixed, from the bottom-right plot in Fig.~\ref{fig:eight} it can be seen that the charge $Q$ increases monotonically as $\phi _{0}$ increases.

The behaviour of $Q$ as a function of mirror radius $r_{m}$ is more complicated, due to the complicated dependence of $r_{m}$ on the parameters $a_{0}$ and $\phi _{0}$ (see Figs.~\ref{fig:five} and \ref{fig:six}).
For smaller fixed $\phi _{0}$, from Fig.~\ref{fig:six} the mirror radius $r_{m}$ increases monotonically as $a_{0}$ decreases, and accordingly we see in the top-left plot in Fig.~\ref{fig:eight} that the charge $Q$ increases monotonically as $r_{m}$ increases.
For larger fixed $\phi _{0}$, it is possible to have two different values of $a_{0}$ giving the same mirror radius $r_{m}$; this is reflected in the curve in the top-left plot in Fig.~\ref{fig:eight} when $\phi _{0}=1.5$ (we anticipate similar behaviour for $\phi _{0}=2.0$ when values of $a_{0}$ above $3.0$ are included).
When $a_{0}$ is fixed, in the bottom-left plot in Fig.~\ref{fig:eight} we see a similar phenomenon. For many values of the mirror radius $r_{m}$, there are two soliton solutions with the same value of $a_{0}$ but two different values of $\phi _{0}$; one of these (with the larger value of $\phi _{0}$) has a larger charge $Q$ than the other.

With fixed mirror radius $r_{m}=18$, in Fig.~\ref{fig:nine} we plot the electric charge $Q$ for those solitons lying on the parameter space curves shown in Fig.~\ref{fig:seven}. The two plots in Fig.~\ref{fig:nine} show the same data, but plotted as a function of the different parameters $\phi _{0}$ and $a_{0}$.  In Fig.~\ref{fig:nine} we see two distinct branches of solutions, corresponding to the two parts of the $r_{m}=18$ contour in the $(a_{0},\phi _{0})$-plane depicted in Fig.~\ref{fig:seven}.  The solutions with larger $\phi _{0}$ for fixed $a_{0}$ have larger charge $Q$ compared to those with smaller $\phi _{0}$.
We dub the branch of solutions with larger charge $Q$ the ``high-charge'' branch and the branch with smaller charge $Q$ the ``low-charge'' branch.

In this paper we consider only the electric charge $Q$ (\ref{eq:charge}) of the solitons and not their mass $M$.  Defining mass as a conserved charge requires first of all a conserved current.  In the conventional approach (for example, to define the usual Komar mass for static asymptotically flat configurations), a conserved current is constructed from either the stress-energy tensor $T_{\mu \nu }$ or Ricci tensor $R_{\mu \nu }$ using the time-like Killing vector $\xi ^{\mu }$, namely either $T_{\mu \nu }\xi ^{\mu }$ or $R_{\mu \nu }\xi ^{\mu }$.  In order to compare the masses of different static spacetimes computed using this conserved current, the time-like Killing vector $\xi ^{\mu }$ must be normalized in a consistent way across all solutions considered.  For static, spherically symmetric, asymptotically flat spacetimes, this is straightforwardly done by insisting that $\xi ^{\mu } \xi _{\mu }\rightarrow -1$ as $r\rightarrow \infty $. The solitons we consider in this paper are static, so each has a time-like Killing vector $\xi ^{\mu }$. However, because we do not have an asymptotically flat spacetime, it is not clear how the normalization of this Killing vector can be consistently chosen to enable meaningful comparisons between different static configurations.  For this reason we have not attempted to define a mass for our soliton solutions.

\section{Stability analysis}
\label{sec:stab}

We now consider time-dependent, spherically symmetric, linear perturbations of the static soliton solutions discussed in the previous section.
The field variables $f$, $h$, $\gamma $, $A_{0}$ and $\Psi $ (\ref{eq:newvars}) now depend on time $t$ as well as the radial coordinate $r$.
We write these in the form:
\begin{align}
f &= \bar{f}(r) + \delta f(t,r), \nonumber \\
h &= \bar{h}(r) + \delta h(t,r), \nonumber \\
\gamma &= \bar{\gamma}(r) + \delta \gamma(t,r), \nonumber \\
A_0 &= \bar{A}_0(r) + \delta A_0(t,r), \nonumber \\
\Psi &= \bar{\psi}(r) + \delta \psi(t,r) ,
\end{align}
where $\bar{f}$ (and similarly for the other variables) denotes the static equilibrium quantity which depends on $r$ only, and $\delta f (t,r)$ is the linear perturbation.
The scalar field perturbation $\delta \psi (t,r)$ is complex; all other quantities are real.
Following \cite{Dolan:2015dha}, we write $\delta \psi (t,r)$ in terms of its real and imaginary parts as follows:
\begin{equation}
\delta \psi (t,r) = \delta u(t,r) + i\delta\dot{w}(t,r),
\label{eq:deltapsi}
\end{equation}
where the imaginary part of $\delta \psi $ is out of phase with the real part.
We note that an arbitrary function of $r$ only can be added to $\delta w$ without changing the scalar field perturbation $\delta \psi $.

The linearized perturbation equations are derived in \cite{Dolan:2015dha} from the dynamical field equations (\ref{eq:dyn}).
It is shown in \cite{Dolan:2015dha} using the perturbed Einstein and Maxwell equations and performing an integration with respect to time that the metric perturbations $\delta f$ and $\delta h$ can be written in terms of the perturbations of the electromagnetic and scalar fields as follows:
\begin{subequations}
\label{eq:metricperts}
\begin{align}
\frac{\df}{\barf} &= \frac{1}{r}\left[\frac{\bpsi}{r}-\bpsi'\right]\delta u - \frac{q\bA\bpsi'}{r}\delta w + \frac{q\bA\bpsi}{r}\delta w' + \delta {\mathcal {F}}(r),
\label{eq:Pt-df} \\
\frac{\delta h}{\bh\sqrt{\bh}} &= -\frac{2q\bg\bpsi'}{r^2\bA'}\delta w + \frac{2q\bg\bpsi}{r^2\bA'}\delta w' + \frac{2}{\sqrt{\bh}\bA'}\delta A'_0
+\delta {\mathcal {H}}(r),
\label{eq:Pt-dh}
\end{align}
\end{subequations}
where $\delta {\mathcal {F}}(r)$ and $\delta {\mathcal {H}}(r)$ are functions of $r$ only, which are arbitrary except that they must satisfy the constraint equation
\begin{widetext}
\begin{align}
\delta {\mathcal {F}}' + \left[\frac{\barf'}{\barf}+\frac{\bh'}{2\bh}+\frac{1}{r}\right]\delta {\mathcal {F}} &= \frac{r\bA\bA'}{2\bg}\delta {\mathcal {H}}' + \frac{r\bA}{2\bg^2}\left[\frac{q^2\bA\sqrt{\bh}\bpsi^2}{r^2} + \frac{\bg\bA'^2}{\bA} + \frac{\barf\bA'\bh'}{2\sqrt{\bh}}\right]\delta {\mathcal {H}},
\label{eq:FHconst}
\end{align}
which has the solution
\begin{equation}
\delta {\mathcal {F}} = \frac {r\bA \bA '}{2\bg }\delta {\mathcal {H}} + \frac {{\mathcal {C}}}{r\bg },
\label{eq:fint}
\end{equation}
where ${\mathcal {C}}$ is an arbitrary constant of integration which is set equal to zero in \cite{Dolan:2015dha}.
The imaginary part of the perturbed scalar field equation can also be integrated with respect to time to give \cite{Dolan:2015dha}
\begin{align}
\label{eq:Pt-KGIm}
0 &= \delta\ddot{w} - \bg^2\delta w'' + \left[- \bg\bg' + \frac{q^2\bpsi^2\bA}{r^2\bA'}\mathcal{A} \right]\delta w' + \left[-q^2\bA^2 -\frac{q^2\bA\bpsi\bpsi'}{r^2\bA'}\mathcal{A} + \frac{\bg\bg'}{r}\right]\delta w  \nonumber \\
&\qquad~~ + q\bA\left[-2 + \frac{\bpsi^2}{r^2} - \frac{\bpsi\bpsi'}{r}\right]\delta u + \frac{q\bA\bpsi}{\bA'}\delta A'_0 - q\bpsi\delta A_0 + \delta {\mathcal {G}}(r),
\end{align}
where we have defined the quantity
\begin{equation}
\mathcal{A}\equiv\barf\bh + r\bA\bA',
\end{equation}
and where $\delta {\mathcal {G}}(r)$ is a function of the radial coordinate $r$ which must satisfy the constraint equation
\begin{align}
\label{eq:dF-dG}
0 &= \delta {\mathcal {F}}' + \left[r\left(\frac{\bar{\psi}}{r}\right)'^2 - \frac{\bA''}{\bA'} - \frac{\bA'}{\bA} - \frac{1}{r} + \frac{\barf'}{\barf}\right]\delta {\mathcal {F}}
+ \frac{q\bA\bpsi}{r\bg^2}\delta {\mathcal {G}} + \frac{\mathcal {C}}{r\bg}\left[\frac{\bA'}{\bA} + \frac{q^2 \bpsi^2 \bA}{\barf r^2 \bA'}\right] .
\end{align}
\end{widetext}
We assume that the perturbations of the metric ($\delta f$, $\delta h$), the electromagnetic potential ($\delta A_{0}$) and the scalar field
($\delta u/r$, $\delta {\dot {w}}/r$) are regular at the origin.
With these assumptions, the functions $\delta {\mathcal {F}}(r)$, $\delta {\mathcal {G}}(r)$, $\delta {\mathcal {H}}(r)$ can be eliminated
from (\ref{eq:metricperts}, \ref{eq:Pt-KGIm}) as follows.
The freedom to add a function of $r$ only to $\delta w $ can be used to set $\delta {\mathcal {H}}(r) \equiv 0$ in (\ref{eq:Pt-dh}) without loss of generality.
Using the expansions (\ref{eq:origin}) and the assumption that $\delta f$ is regular at the origin, it follows from (\ref{eq:Pt-df}) that $\delta {\mathcal {F}}$ must also be finite at the origin.
Therefore, setting $\delta {\mathcal {H}}=0$ in (\ref{eq:fint}), the only possibility is ${\mathcal {C}}=0$ and therefore $\delta {\mathcal {F}} \equiv 0$.
Finally, from (\ref{eq:dF-dG}), we have that $\delta {\mathcal {G}} \equiv 0$ as well.

The perturbation equation (\ref{eq:Pt-KGIm}) therefore simplifies.
The remaining perturbation equations (comprising the real part of the perturbed scalar field equation and one of the Einstein field equations) also simplify (their derivation can be found in \cite{Dolan:2015dha}).
Altogether we have three linearized perturbation equations:
\begin{widetext}
\begin{subequations}
\begin{align}\label{eq:Pt-Real}
0 &= \delta\ddot{u} - \bg^2\delta u'' -\bg\bg'\delta u' + \left[3q^2\bA^2+\frac{\bg\bg'}{r}-\barf\bh \left(\frac{\bar{\psi}}{r}\right)'^2+\frac{\barf\bA'^2}{2}\left(\left(\frac{\bar{\psi}}{r}\right)^2+\bpsi'^2\right)-\frac{\barf\bpsi\bpsi'\bA'^2}{r}\right]\delta u + 2q\bA\bg^2\delta w'' \nonumber \\
&\quad + q\barf\bA\left[2\sqrt{\bh}\bg'+\left(-\frac{\bA'}{\bA}\mathcal{A}+\frac{\bh}{r}+\frac{r\bA'^2}{2}\right)\left(\frac{\bar{\psi}}{r}\right)'\bpsi\right]\delta w' \nonumber \\
& \quad + q\bA\left[2q^2\bA^2 -\frac{2\bg\bg'}{r}+\bg\bpsi'\left(\frac{\bar{\psi}}{r}\right)'\left(\frac{\bg\bA'}{\bA}-\bg'-\frac{\bg}{r}\right)\right]\delta w , \\ \label{eq:Pt-Im}
0 &= \delta\ddot{w} - \bg^2\delta w'' + \left[-\bg\bg' + \frac{q^2\bA\bpsi^2}{r^2\bA'}\mathcal{A}\right]\delta w' + \left[-q^2\bA^2 - \frac{q^2\bA\bpsi\bpsi'}{r^2\bA'}\mathcal{A}+\frac{\bg\bg'}{r}\right]\delta w -q\bA\left[2+\bar{\psi}\left(\frac{\bar{\psi}}{r}\right)'\right]\delta u \nonumber \\
&\qquad + \frac{q\bA\bpsi}{\bA'}\delta A'_0 - q\bpsi\delta A_0 ,
\\
\label{eq:Pt-Const}
0 &= \frac{q\bpsi}{\bA' r^2}\mathcal{A}\delta w'' + \frac{q\bpsi\bA}{r^2}\left[\frac{\bg'}{\bA\bA'\bg}\mathcal{A} - \frac{q^2\bpsi^2\bh}{r^2\bA'^2}\right]\delta w' + \frac{q\bpsi\bA}{r^2}\left[\frac{\mathcal{A}}{r\bA\bA'\bg}\left(-\bg'+\frac{rq^2\bA^2}{\bg}\right) + \frac{q^2\bh\bpsi\bpsi'}{r^2\bA'^2}\right]\delta w \nonumber \\
&\qquad -\left(\frac{\bar{\psi}}{r}\right)'\delta u' -\left[\left(\frac{\bar{\psi}}{r}\right)''+\left(\frac{1}{r}+\frac{\bg'}{\bg}\right)\left(\frac{\bar{\psi}}{r}\right)'\right]\delta u + \left[\frac{\delta A'_0}{\bA'}\right]' .
\end{align}
\label{eq:perteqnsfinal}
\end{subequations}
\end{widetext}

We consider time-periodic perturbations of the form
\begin{align}
\delta u (t,r) & = \text{Re}\left[ e^{-i\sigma t}\tilde{u}(r) \right],
\nonumber \\
\delta w (t,r)  & = \text{Re}\left[ e^{-i\sigma t}\tilde{w}(r) \right],
\nonumber \\
\delta A_0(t,r) & = \text{Re} \left[ e^{-i\sigma t}\tilde{A}_{0}(r) \right] ,
\label{eq:timep}
\end{align}
where ${\tilde {u}}$, ${\tilde {w}}$ and ${\tilde {A}}_{0}$ are complex functions of $r$ only.
Near the origin, we assume that the functions of $r$ in (\ref{eq:timep}) have the following expansions:
\begin{align}
{\tilde {u}} & = r \sum _{j=0}^{\infty } u_{j}r^{j},
\nonumber \\
{\tilde {w}} & = r\sum _{j=0}^{\infty } w_{j}r^{j},
\nonumber \\
{\tilde {A}}_{0} & = \sum _{j=0}^{\infty } \alpha _{j}r^{j},
\label{eq:originperts}
\end{align}
where the $u_j$, $w_j$ and $\alpha _{j}$ are constants, so that the perturbations of the scalar and electromagnetic fields are regular at the origin.

Substituting the expansions (\ref{eq:originperts}) into the perturbation equations (\ref{eq:perteqnsfinal}) and comparing powers of $r$, we find that
\begin{equation}
u_{1}= w_{1} = \alpha _{1} = 0 = u_{3} = w_{3} = \alpha _{3},
\end{equation}
where we have also used the fact that the perturbed Ricci scalar curvature must be finite at the origin.
We also find that $\alpha _{2}$ and $u_{2}$ are given in terms of $\sigma ^{2}$, $\alpha _{0}$, $u_{0}$, $w_{0}$ and $w_{2}$.
Subsequent terms in the expansions (\ref{eq:originperts}) are also given in terms of the five quantities $\sigma ^{2}$, $\alpha _{0}$, $u_{0}$,
$w_{0}$ and $w_{2}$.

At the mirror $r=r_m$, the scalar field perturbation $\delta \psi $ must vanish for all $t$, so we require
\begin{equation}
{\tilde {u}}(r_m)= 0 = {\tilde {w}}(r_m).
\label{eq:rmconst}
\end{equation}
The values of the metric and electromagnetic field perturbations are unconstrained at $r=r_{m}$.

The boundary conditions (\ref{eq:rmconst})  give only two constraints on the field perturbations.  We therefore expect to have just two free parameters in the expansions (\ref{eq:originperts}) in order to obtain a spectrum of eigenvalues $\sigma ^{2}$.
At the moment we have five free parameters,  $\sigma ^{2}$, $\alpha _{0}$, $u_{0}$, $w_{0}$ and $w_{2}$.
Since we have linear perturbation equations (\ref{eq:perteqnsfinal}), we have freedom to set the overall scale of the perturbations. We choose to fix $u_{0}=0.5$, leaving four arbitrary parameters. Of these, only two are gauge-invariant.

Performing an infinitesimal $U(1)$ gauge transformation (\ref{eq:gauge-transform}) with $\chi = {\text {Re}} [ \chi _{0} e^{-i\sigma t}] $ where $\chi _{0}$ is a complex constant gives
\begin{equation}
\sigma  {\tilde {w}} \rightarrow  \sigma {\tilde {w}} +  i \bpsi \chi _{0} ,
\qquad
{\tilde {A}}_{0} \rightarrow {\tilde {A}}_{0} - i q^{-1} \sigma \chi _{0}.
\end{equation}
Bearing in mind that ${\tilde {u}}$, ${\tilde {w}}$ and ${\tilde {A}}_{0}$ are complex functions, we can therefore choose $\chi _{0}$ to be (using the expansions (\ref{eq:origin}) near the origin)
\begin{equation}
\chi _{0} = \frac { iw_{0} \sigma }{\phi _{0}},
\end{equation}
and hence set $w_{0}=0$ without loss of generality.

There is also a residual diffeomorphism freedom, corresponding to a redefinition of the time coordinate (this is the time-dependent analogue of the rescaling (\ref{eq:newvars2})).  Under an infinitesimal coordinate transformation generated by the vector
\begin{equation}
V = \left( {\text {Re}} [x_{0}e^{-i\sigma t}] , 0, 0,  0 \right) ,
\end{equation}
where $x_{0}$ is a complex constant,
the scalar field perturbations and metric perturbation $\delta f$ are unchanged.
The electromagnetic potential perturbation $\delta A_{0}$ and metric perturbation $\delta h$ transform in such a way that the relation (\ref{eq:Pt-dh}) is unchanged by this coordinate transformation.
In particular, the electromagnetic potential perturbation $\delta A_{0}$ transforms as follows (see, for example, the discussion in \cite{Nolan:2015vca}):
\begin{equation}
{\tilde {A}}_{0} \rightarrow {\tilde {A}}_{0} -i \sigma \bA x_{0} .
\end{equation}
Therefore, choosing (again using the expansions (\ref{eq:origin}) near the origin)
\begin{equation}
x_{0} =  -\frac {i\alpha _{0}}{\sigma a_{0}},
\end{equation}
we may set $\alpha _{0}=0$ without loss of generality.
We are now left with two free parameters, namely $\sigma ^{2}$ and $w_{2}$.
We note that in the black hole case \cite{Dolan:2015dha}, imposing ingoing boundary conditions at the event horizon fixes both the residual $U(1)$ gauge freedom and the diffeomorphism freedom that we have here for soliton solutions.

We can now integrate the perturbation equations (\ref{eq:perteqnsfinal}) numerically. We use the expansions (\ref{eq:originperts}) as initial conditions close to the origin, and seek values of the shooting parameters $\sigma ^{2}$, $w_{2}$ such that the boundary conditions on the mirror (\ref{eq:rmconst}) are satisfied.
The perturbation equations (\ref{eq:perteqnsfinal}) and boundary conditions (\ref{eq:originperts}) depend only on $\sigma ^{2}$ since we are considering time-periodic perturbations and the only time derivatives in the perturbation equations are $\delta {\ddot {u}}$ and $\delta {\ddot {w}}$.  The perturbation equations and boundary conditions therefore define an eigenvalue problem for $\sigma ^{2}$.
When $\sigma ^{2}$ is real, we can consider real perturbation functions ${\tilde {u}}$, ${\tilde {w}}$ and ${\tilde {A}}_{0}$ without loss of generality.
The system of perturbation equations (\ref{eq:perteqnsfinal}) can be written in the form
\begin{equation}
\sigma ^{2} {\mathfrak {M}} \left(
\begin{array}{c}
{\tilde {u}} \\
{\tilde {v}} \\
{\tilde {A}}_{0}
\end{array}
\right)
= {\mathfrak {O}} \left(
\begin{array}{c}
{\tilde {u}} \\
{\tilde {v}} \\
{\tilde {A}}_{0}
\end{array}
\right) ,
\end{equation}
where ${\mathfrak {O}}$ is a second-order differential operator and ${\mathfrak {M}}={\text {Diag}} \{ 1, 1, 0 \} $.
The operator ${\mathfrak {O}}$ is not symmetric and therefore the eigenvalue $\sigma ^{2}$ is not necessarily real, although we did not find any complex eigenvalues $\sigma ^{2}$ for all the solitons we investigated with the mirror at the first zero of the equilibrium scalar field.

Our particular interest is in the sign of the imaginary part of the mode frequency, ${\text {Im}} (\sigma )$.
If ${\text {Im}}(\sigma )>0$, then the perturbations (\ref{eq:timep}) are exponentially growing in time and the corresponding static configuration is unstable.
If ${\text {Im}} (\sigma )\le 0$, then the perturbations do not grow with time and the corresponding static configuration is stable.
In our numerical analysis below, we find that $\sigma ^{2}$ is real.  Therefore, if $\sigma ^{2}>0$, the frequency $\sigma $ is also real and the solitons are stable. However, if $\sigma ^{2}<0$, then the frequency $\sigma $ is purely imaginary and there will be perturbations which grow exponentially with time.
In this case the solitons are unstable.

With the scalar field charge $q$ fixed to be $0.1$ using the scaling symmetry (\ref{eq:newvars1}), we have a two-parameter $(a_{0}, \phi _{0})$ space of static equilibrium solutions, shown in Fig.~\ref{fig:four}.
We now present a selection of numerical results exploring perturbations of static charged-scalar solitons in this phase space.
For each static equilibrium solution, we search for values of $\sigma ^{2}$ and $w_{2}$ such that the resulting perturbations satisfy the boundary conditions (\ref{eq:originperts}, \ref{eq:rmconst}).  In our plots we show the lowest value of $\sigma ^{2}$ found by this method.
Throughout our investigation, the mirror is located at the first zero of the equilibrium scalar field.

\begin{figure}
\includegraphics[width=8.5cm]{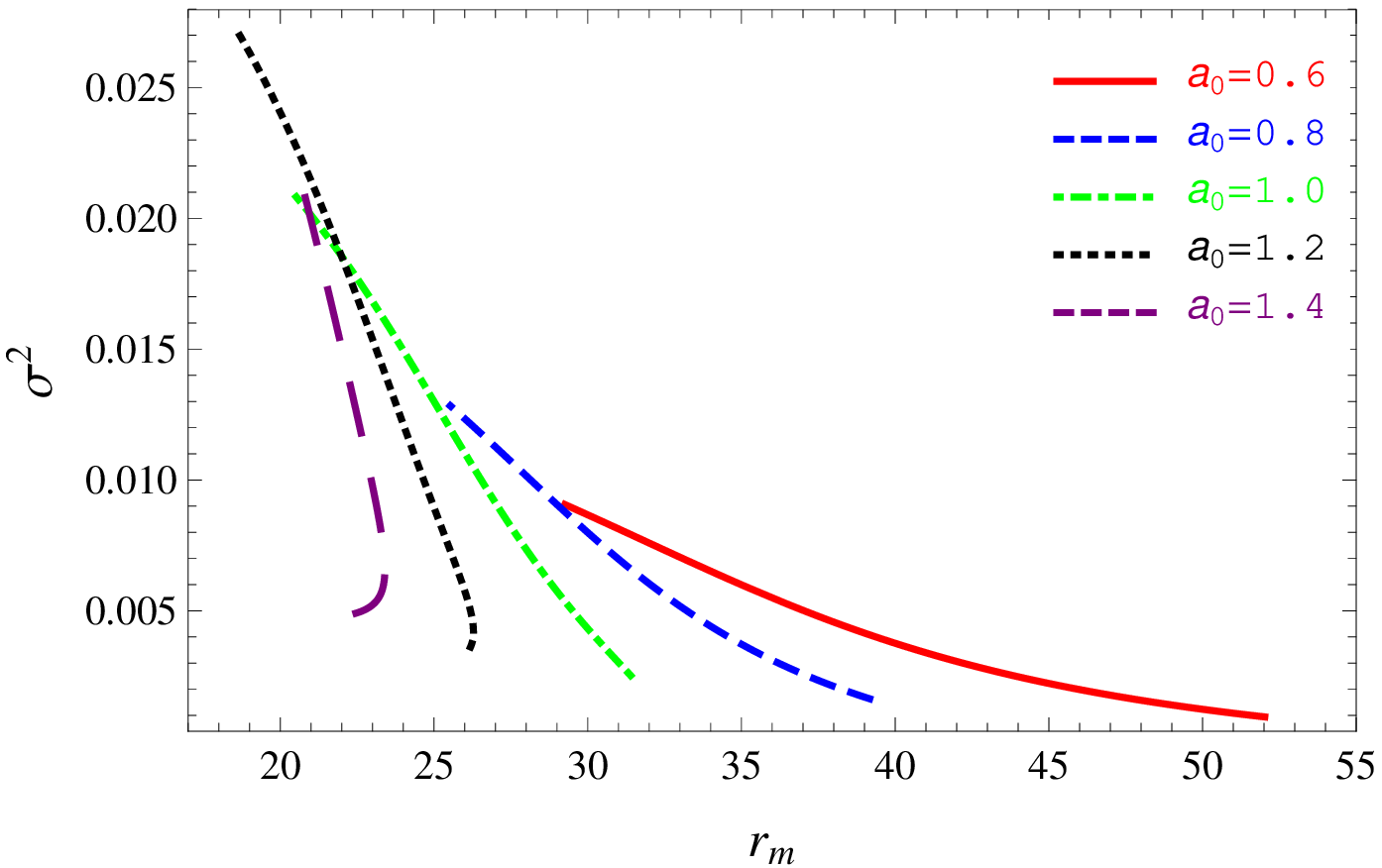}
\includegraphics[width=8.5cm]{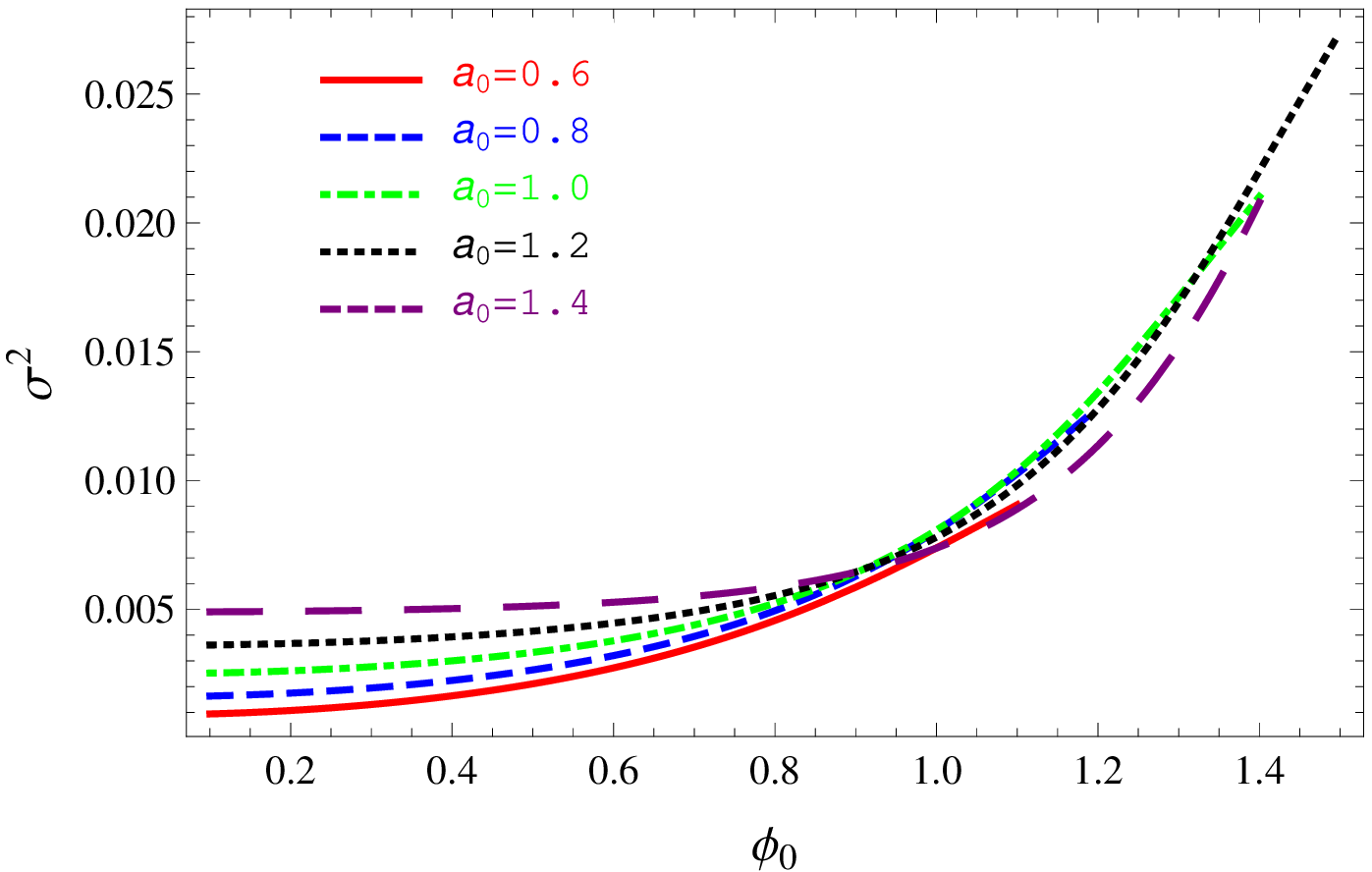}
\caption{Smallest eigenvalue $\sigma ^{2}$ for scalar field charge $q=0.1$, various fixed values of $a_{0}$ and $\phi _{0}\in \left( 0.1, 1.6 \right)$.
Top: $\sigma ^{2}$ as a function of the mirror radius $r_{m}$. Bottom: the same data for $\sigma ^{2}$, but plotted as a function of $\phi _{0}$.}
\label{fig:ten}
\end{figure}

In Fig.~\ref{fig:ten} we plot the smallest eigenvalue $\sigma ^{2}$ for fixed scalar field charge $q=0.1$, various values of $a_{0}$ and values of $\phi _{0}$ in the interval $(0.1, 1.6)$. For values of $\phi _{0}$ larger than $1.6$ there are no static equilibrium solutions for the values of $a_{0}$ shown; our numerical method breaks down when $\phi _{0}$ is very small and the mirror is far from the origin.
In the two plots in Fig.~\ref{fig:ten} we show the same data for $\sigma ^{2}$, firstly as a function of the mirror radius $r_{m}$ and secondly as a function of $\phi _{0}$.
As in Fig.~\ref{fig:five}, when $a_{0}=1.4$ there are equilibrium soliton solutions with different values of $\phi _{0}$ but the same mirror radius $r_{m}$.
This is why the $a_{0}=1.4$ curve for $\sigma ^{2}$ as a function of $r_{m}$ is double-valued.
In general we find that $\sigma ^{2}$ decreases as $r_{m}$ increases and $\phi _{0}$ decreases for fixed $a_{0}$.
For all values of $a_{0}$ and $\phi _{0}$ considered in Fig.~\ref{fig:ten}, the lowest value of $\sigma ^{2}$ that we find is positive. Therefore,
all equilibrium solutions with these values of $a_{0}$ and $\phi _{0}$ are stable.

\begin{figure*}
\includegraphics[width=8.5cm]{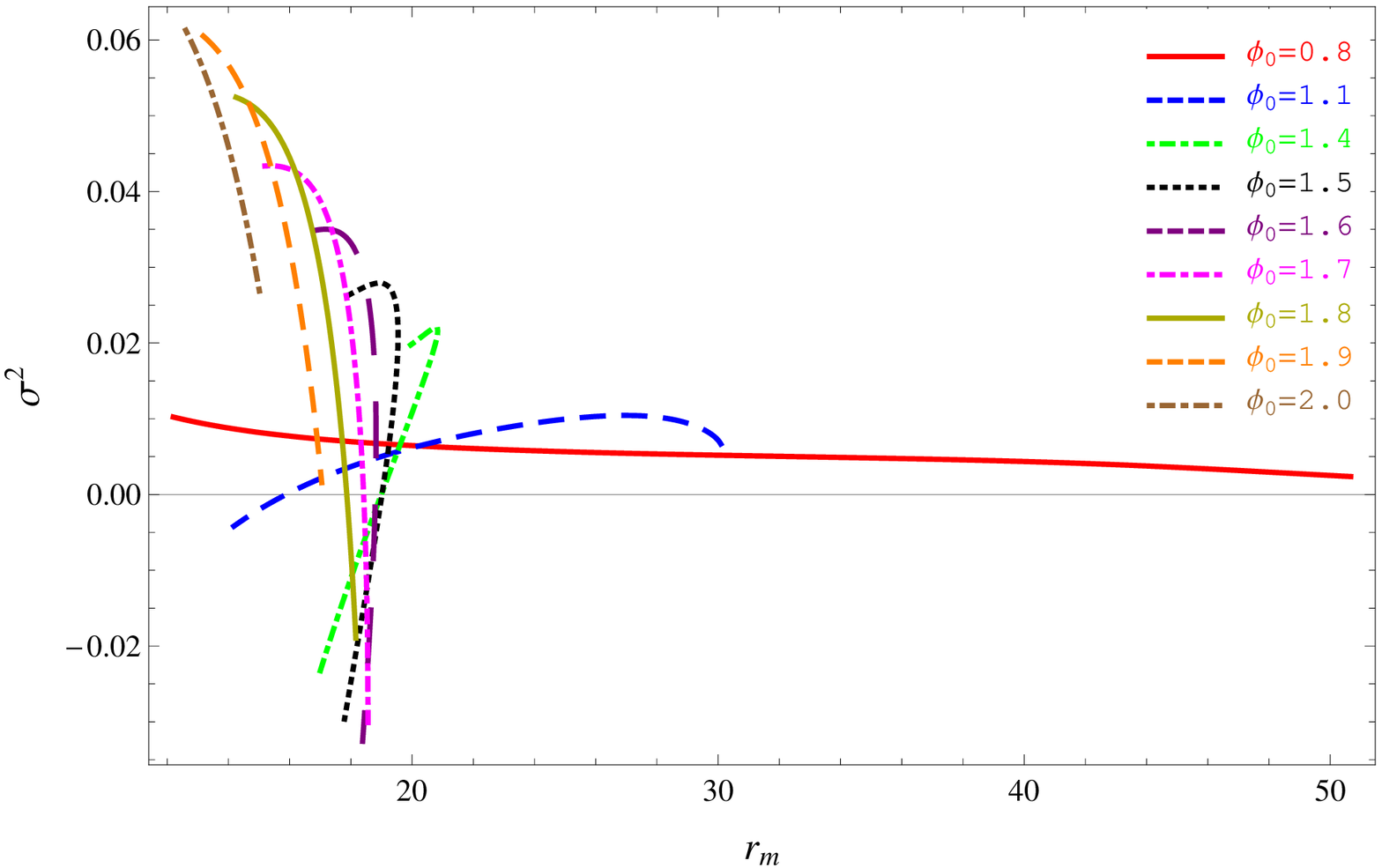}
\includegraphics[width=8.5cm]{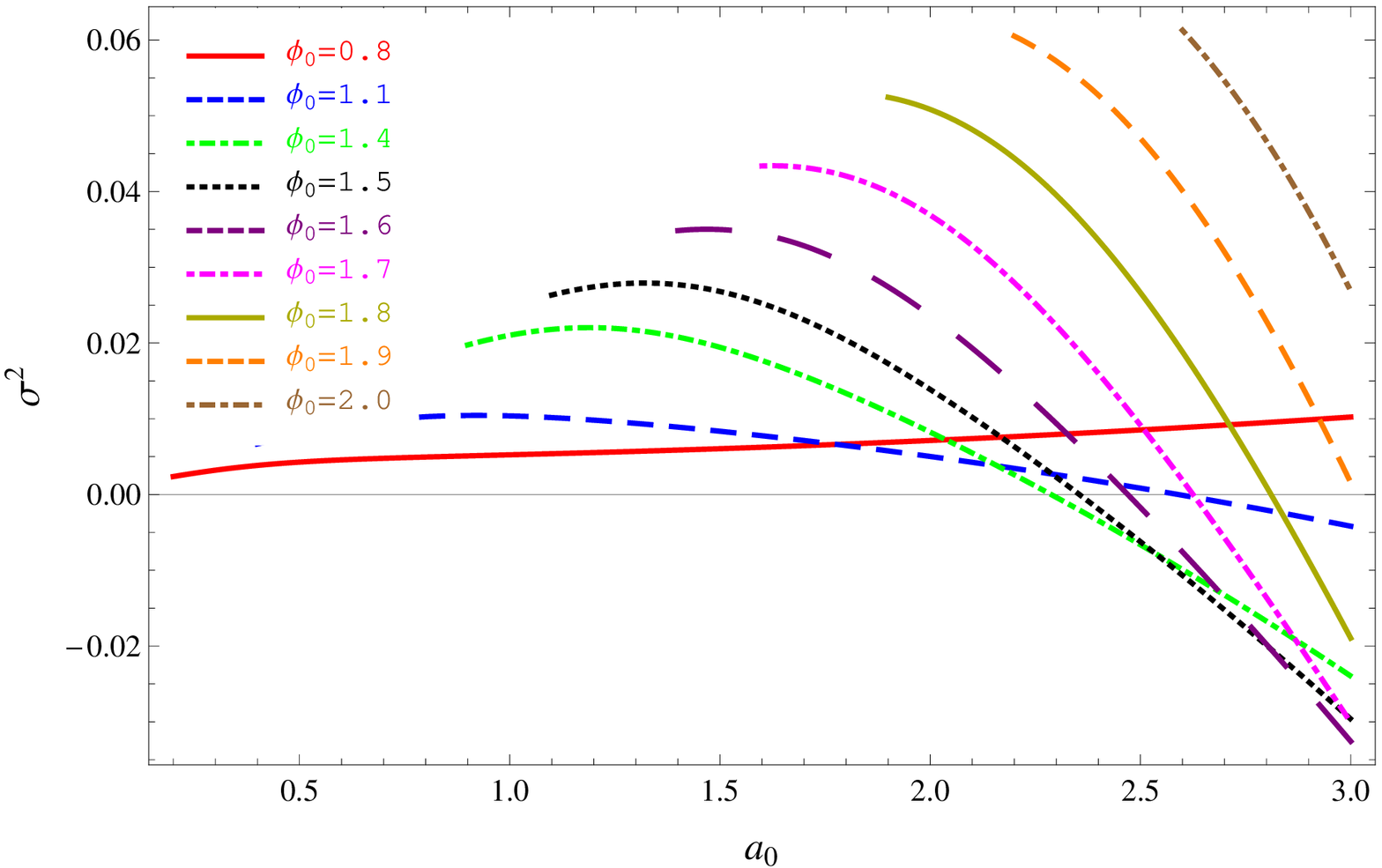}
\includegraphics[width=8.5cm]{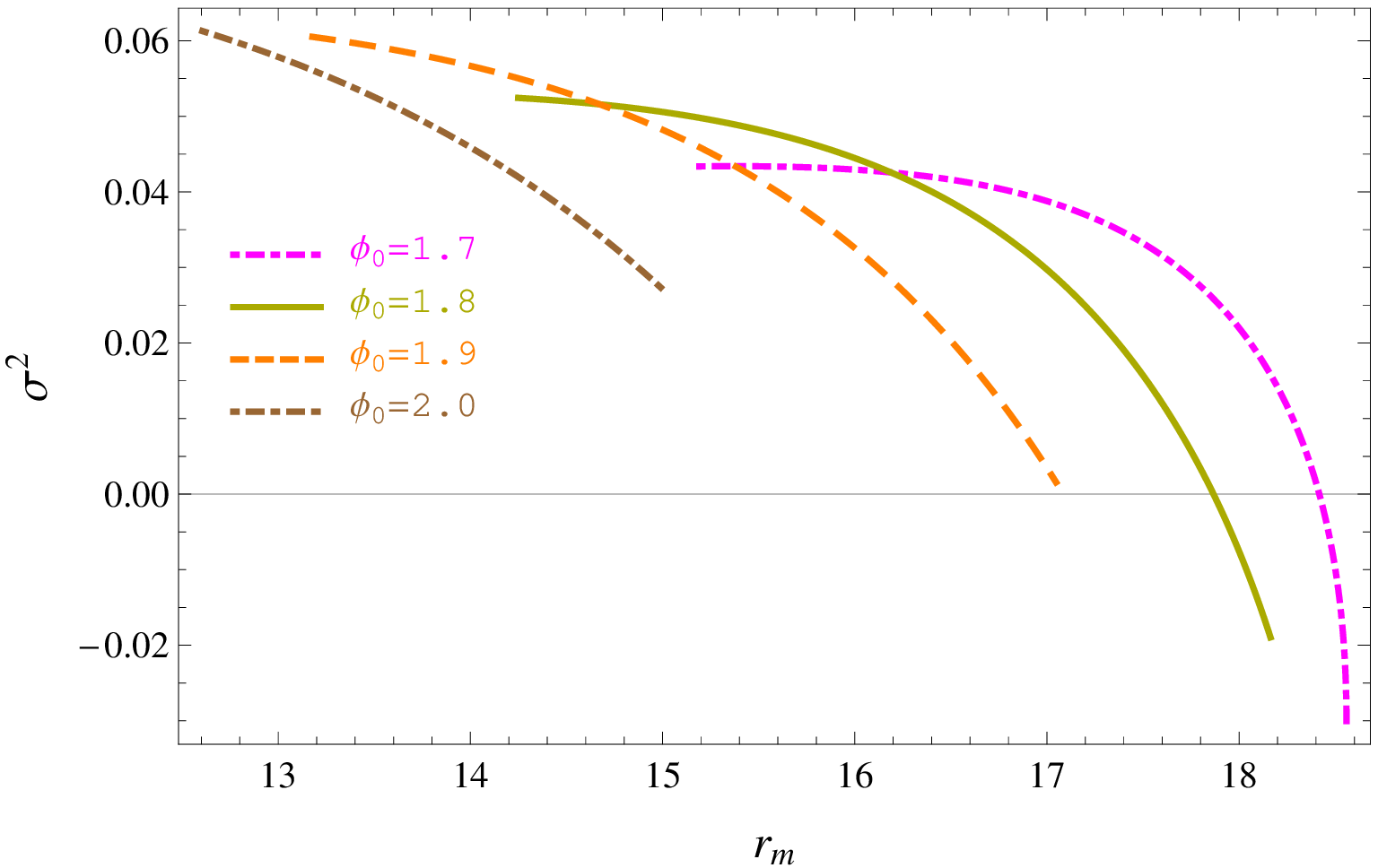}
\includegraphics[width=8.5cm]{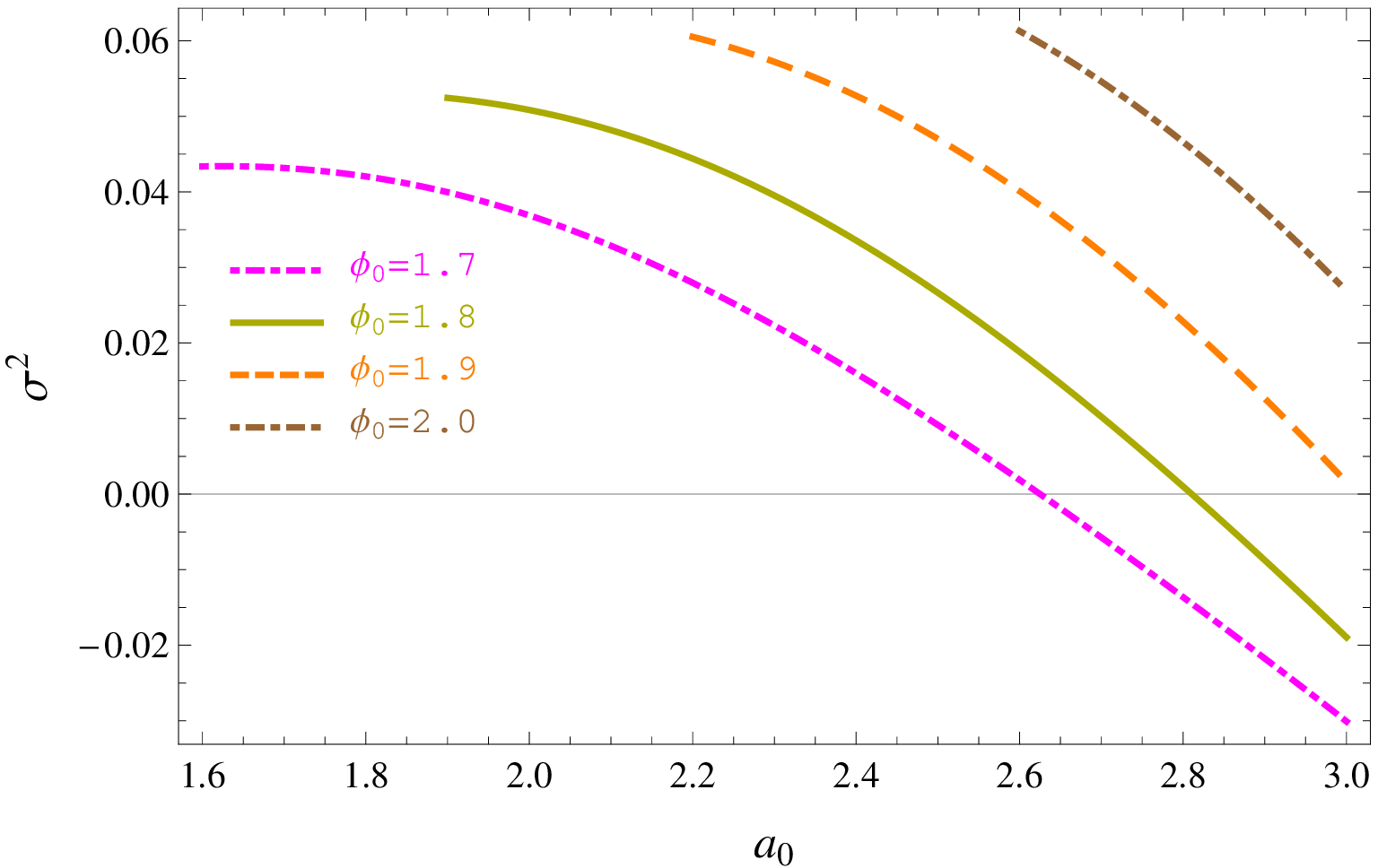}
\includegraphics[width=8.5cm]{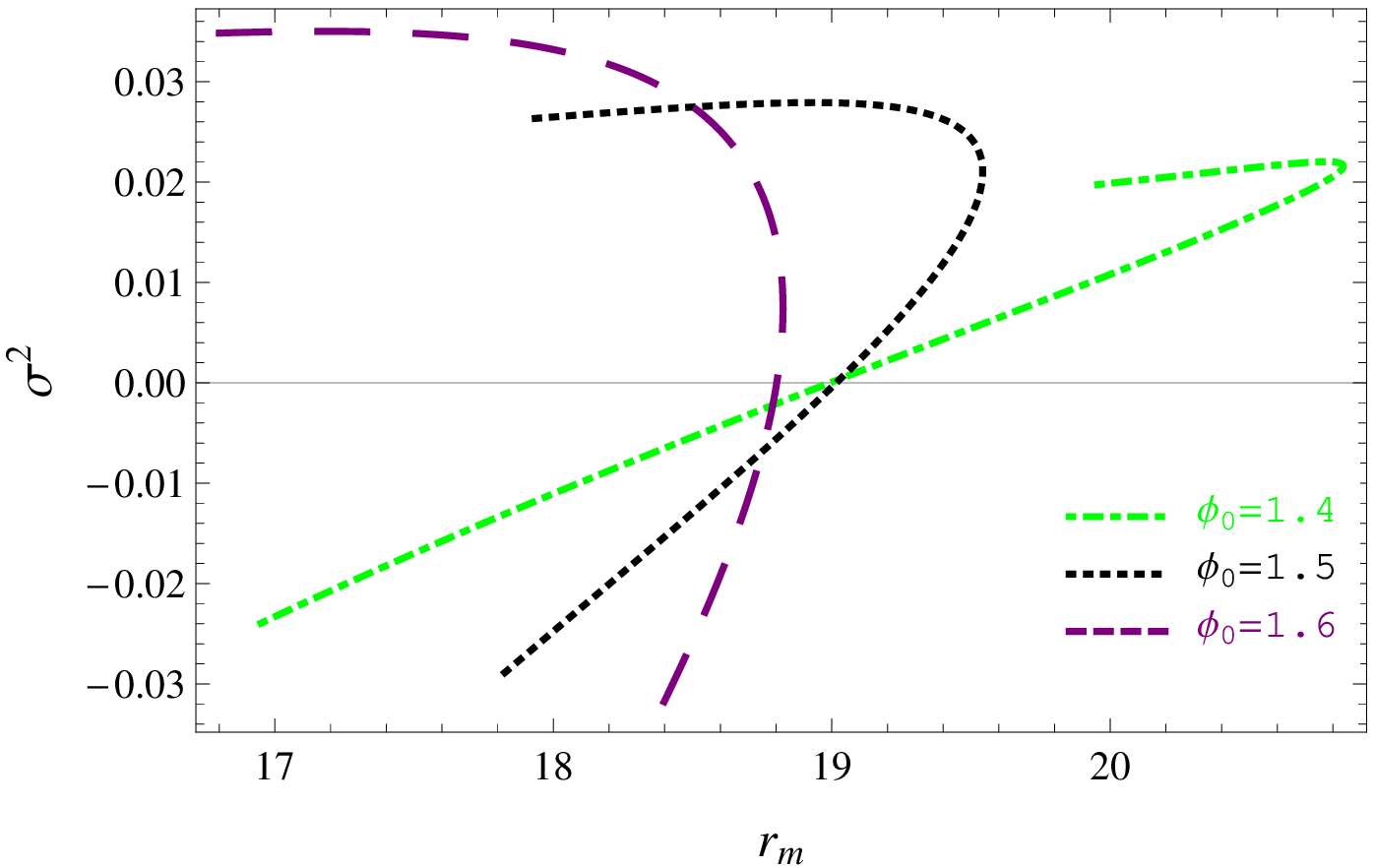}
\includegraphics[width=8.5cm]{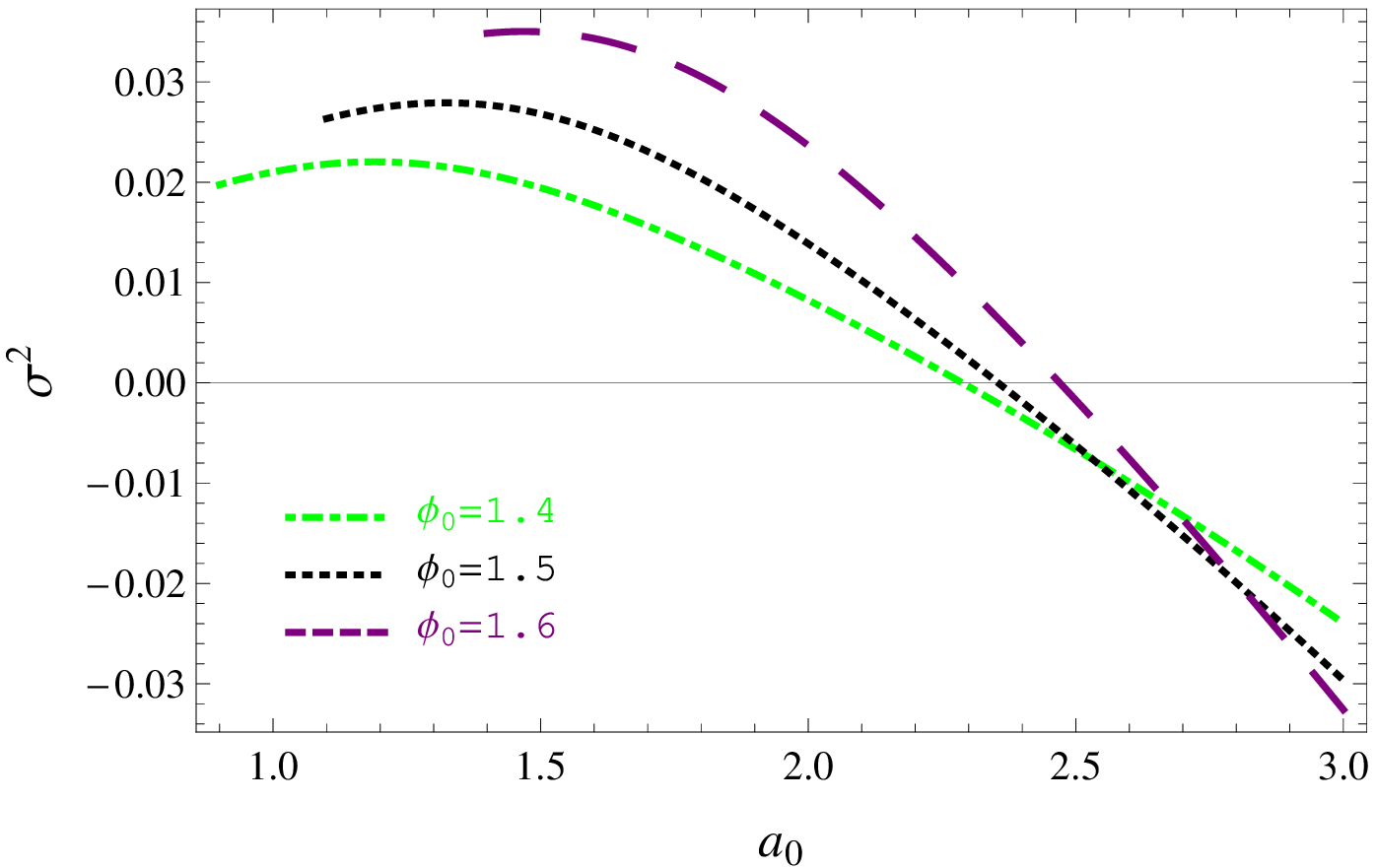}
\includegraphics[width=8.5cm]{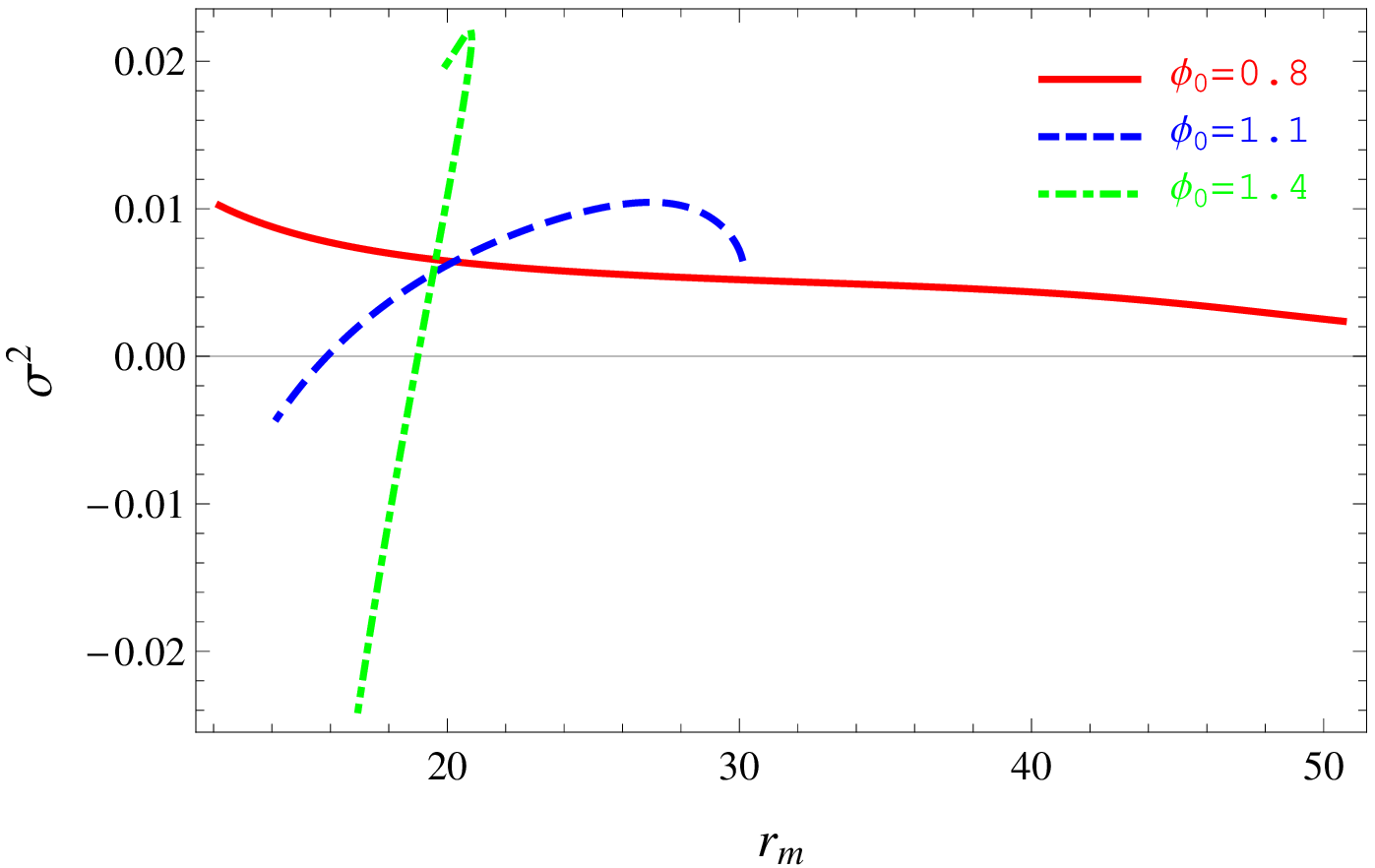}
\includegraphics[width=8.5cm]{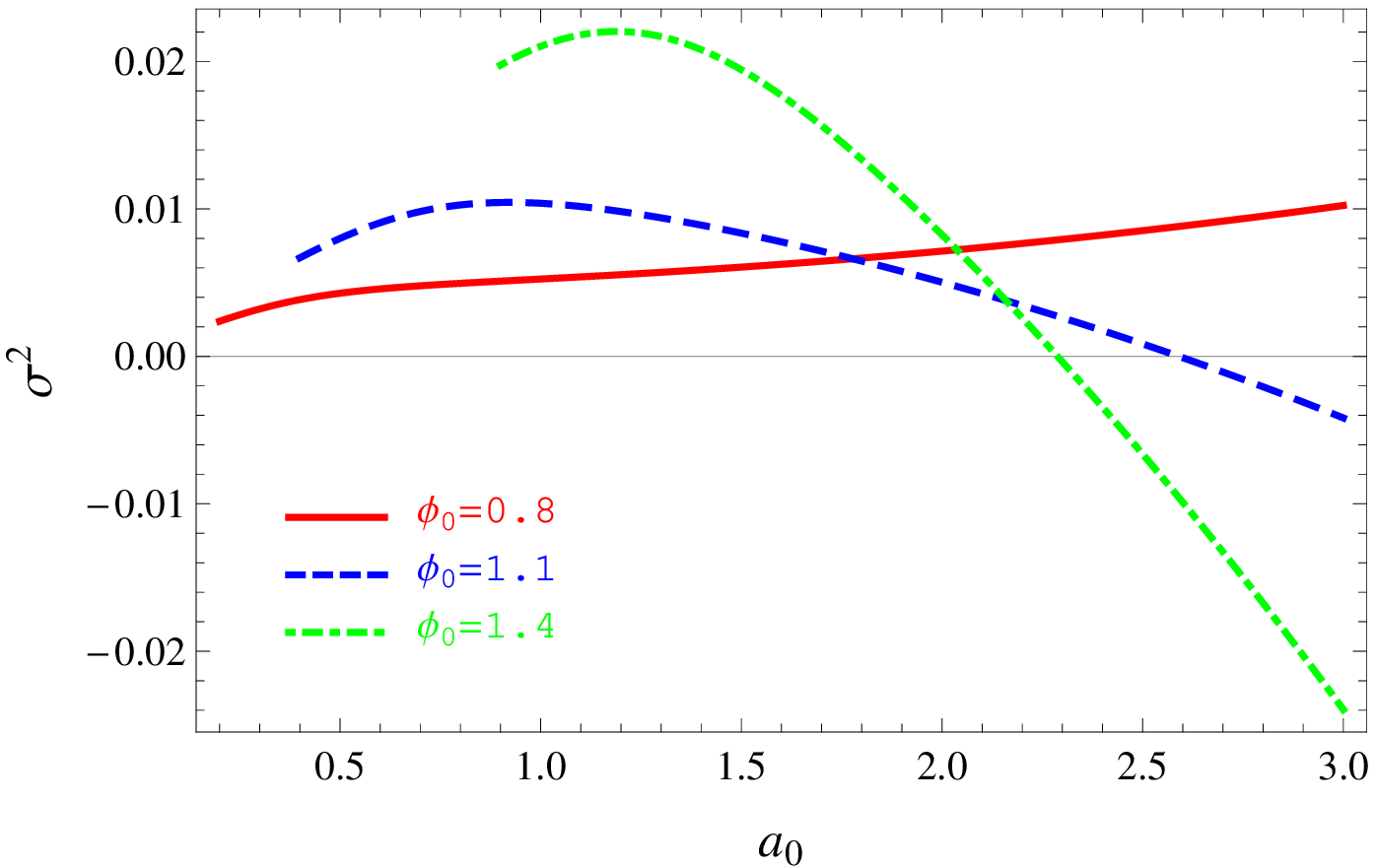}
\caption{Smallest eigenvalue $\sigma ^{2}$ for scalar field charge $q=0.1$, various fixed values of $\phi _{0}$ and $a_{0}\in \left( 0.2, 3.0 \right)$.
Left-hand plots: $\sigma ^{2}$ as a function of the mirror radius $r_{m}$. Right-hand plots: the same data for $\sigma ^{2}$, but plotted as a function of $a_{0}$.
To make the behaviour more visible, the data in the top row of plots is repeated in the remaining plots, just for a few values of $\phi _{0}$.}
\label{fig:eleven}
\end{figure*}

We explore the parameter space of equilibrium solutions further in Fig.~\ref{fig:eleven}.  Here we fix the scalar field charge to be $q=0.1$, consider various fixed values of $\phi _{0}$ and then vary $a_{0}$ in the interval $\left( 0.2, 3 \right)$.
We focus particularly on larger values of $\phi _{0}$ and $a_{0}$.  The plots on the left-hand-side of Fig.~\ref{fig:eleven} show the lowest value of $\sigma ^{2}$ as a function of the mirror radius $r_{m}$; the plots on the right-hand-side show the same data for $\sigma ^{2}$, but as a function of $a_{0}$.
In the top row in Fig.~\ref{fig:eleven} we show data for various fixed values of $\phi _{0} \in [0.8, 2.0]$.  The same data is shown in the lower plots in Fig.~\ref{fig:eleven}, but in each case for a small number of fixed values of $\phi _{0}$, in order to make the behaviour of $\sigma ^{2}$ easier to see.
The corresponding plots of the mirror radius $r_{m}$ as a function of $a_{0}$ for the same values of $\phi _{0}$ can be found in Fig.~\ref{fig:six}.

For the smallest value of $\phi _{0}$ considered in Fig.~\ref{fig:eleven}, namely $\phi _{0}=0.8$, we see that $\sigma ^{2}$ is always positive and increases as $a_{0}$ increases and $r_{m}$ decreases.  However, the behaviour for larger values of $\phi _{0}$ is markedly different.  For all $\phi _{0}\ge 1.1$ shown in Fig.~\ref{fig:eleven}, we see that the lowest value of $\sigma ^{2}$ decreases as $a_{0}$ increases, and becomes negative for sufficiently large $a_{0}$.
We deduce that the solitons with smaller values of $a_{0}$ are stable, but those for larger $a_{0}$ are unstable.
The value of $a_{0}$ at which $\sigma ^{2}$ passes through zero shows complicated behaviour:  at first it decreases as $\phi _{0}$ increases, but for $\phi _{0} \ge 1.4$ it increases as $\phi _{0}$ increases.

The behaviour of the lowest value of $\sigma ^{2}$ as a function of the mirror radius $r_{m}$ can be seen in the left-hand plots in Fig.~\ref{fig:eleven}, and is also quite complicated. For some values of $r_{m}$ it is double-valued because there are two values of $a_{0}$ for that particular $\phi _{0}$ for which the mirror has the same radius $r_{m}$, see Fig.~\ref{fig:six}.
For $\phi _{0}=0.8$, the mirror radius $r_{m}$ decreases monotonically as $a_{0}$ increases, and, as already noted, $\sigma ^{2}$ is positive for all values of $r_{m}$ studied.
For larger values of $\phi _{0}$ in Fig.~\ref{fig:eleven}, the lowest value of $\sigma ^{2}$ is negative for some values of $r_{m}$. For some small $r_{m}$ there are two values of $\sigma ^{2}$, these correspond to different values of $a_{0}$, with the smaller values of $\sigma ^{2}$ arising for larger values of $a_{0}$.
For all the cases we have examined, the negative values of $\sigma ^{2}$ arise when $r_{m}$ is less than about 20.
In Fig.~\ref{fig:eleven} we have studied values of $a_{0}$ only up to 3.
For this range of values of $a_{0}$, it can be seen that when $\phi _{0} = 1.9$ or $2.0$ that $\sigma ^{2}$ is always positive. However, we expect that $\sigma ^{2}$ will become negative if we consider larger values of $a_{0}$.

From this analysis we conclude that the stability of the soliton solutions depends on the values of the scalar field $\phi _{0}$ and electromagnetic potential $a_{0}$ at the origin.  Roughly speaking, when these are both small (and the mirror radius $r_{m}$ is large) the solitons appear to be stable; we were unable to find any negative values of the eigenvalue $\sigma ^{2}$.
However, for sufficiently large $\phi _{0}$ and $a_{0}$  (and, consequently, sufficiently small mirror radius $r_{m}$), we find negative values of $\sigma ^{2}$ and some of the solitons are unstable.

\begin{figure}
\includegraphics[width=8.5cm]{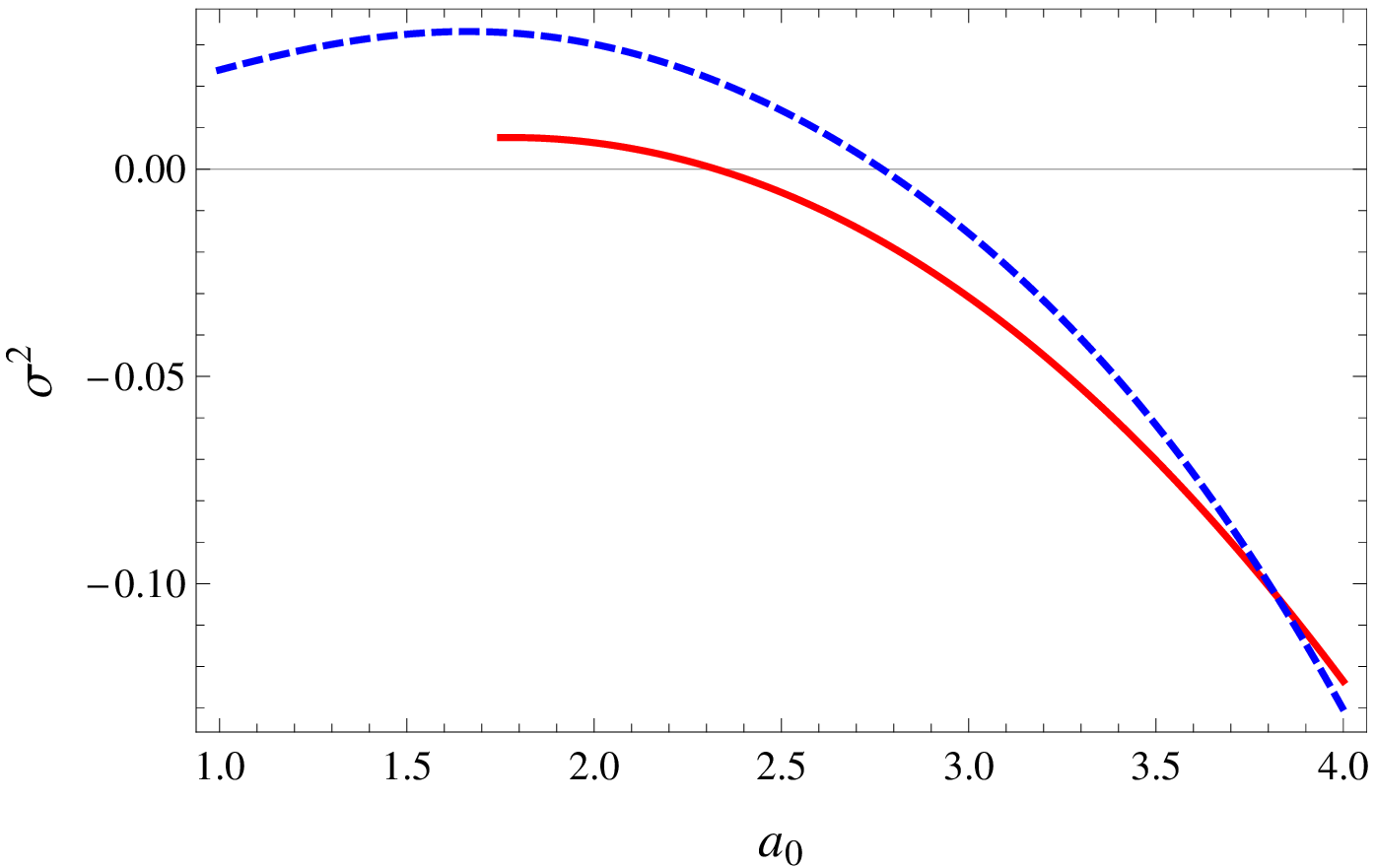}
\includegraphics[width=8.5cm]{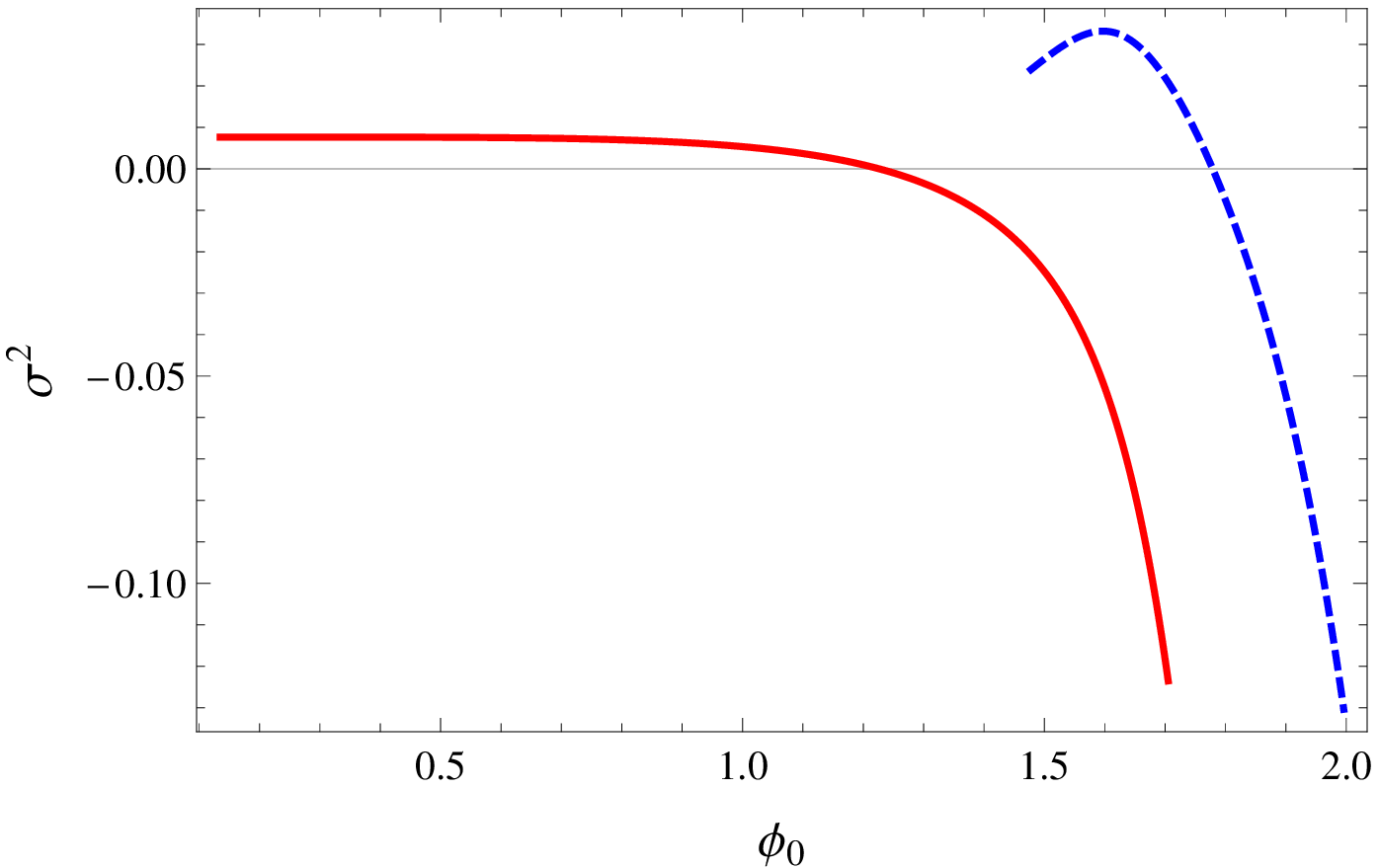}
\caption{Smallest eigenvalue $\sigma ^{2}$ for solitons with fixed mirror radius $r_{m}=18$ and scalar charge $q=0.1$. The equilibrium solutions considered are those lying on the portions of the $r_{m}=18$ contour in the $(a_{0},\phi _{0})$-plane shown in Fig.~\ref{fig:seven}. The same data is shown in the two plots. Top: $\sigma ^{2}$ as a function of $a_{0}$. Bottom: $\sigma ^{2}$ as a function of $\phi _{0}$.}
\label{fig:twelve}
\end{figure}

\begin{figure}
\includegraphics[width=8.5cm]{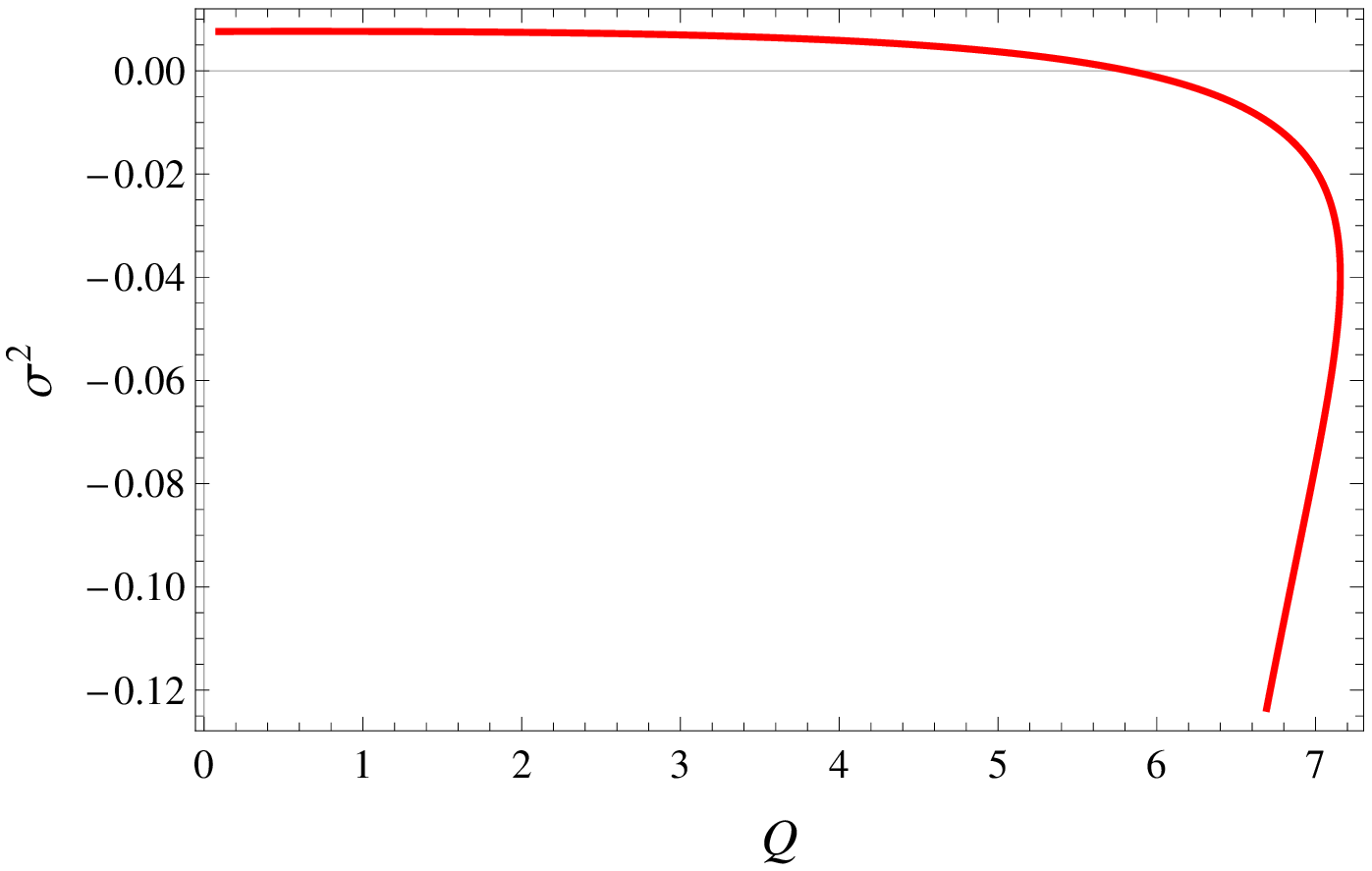}
\includegraphics[width=8.5cm]{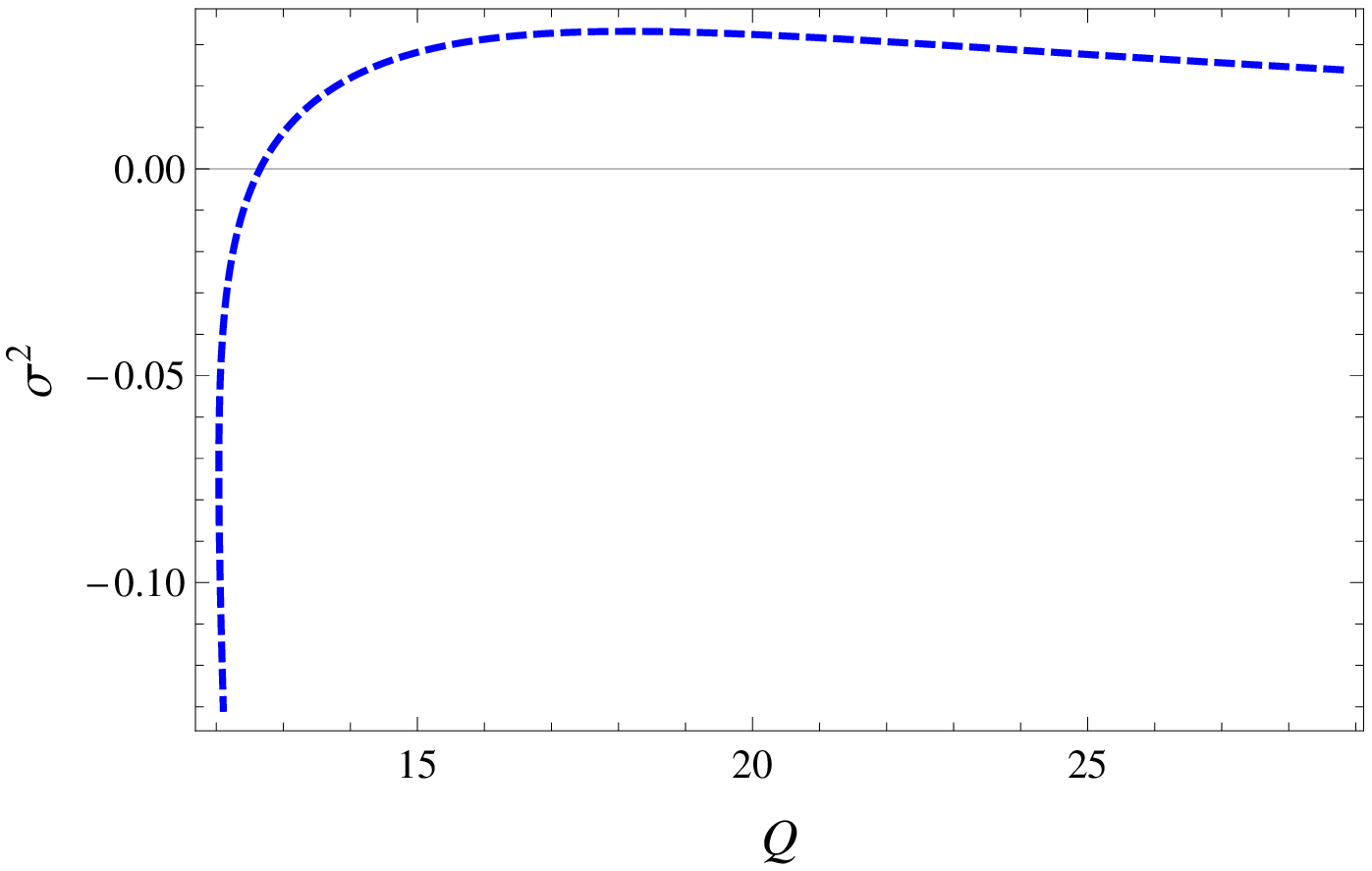}
\caption{Smallest eigenvalue $\sigma ^{2}$ for solitons with fixed mirror radius $r_{m}=18$ and scalar charge $q=0.1$, plotted as a function of the soliton electric charge $Q$ (\ref{eq:charge}). The equilibrium solutions considered are those lying on the portions of the $r_{m}=18$ contour in the $(a_{0},\phi _{0})$-plane shown in Fig.~\ref{fig:seven}. The same data as in Fig.~\ref{fig:twelve} is plotted. Top: $\sigma ^{2}$ as a function of $Q$ for the ``low charge'' branch (the branch of solutions with smaller $\phi _{0}$ for fixed $a_{0}$). Bottom: $\sigma ^{2}$ as a function of $Q$ for the ``high charge'' branch (the branch of solutions with larger $\phi _{0}$ for fixed $a_{0}$).}
\label{fig:thirteen}
\end{figure}

To see how the stability of the solitons depends on the parameters $a_{0}$ and $\phi _{0}$ when the mirror radius $r_{m}$ is fixed, in Figs.~\ref{fig:twelve} and \ref{fig:thirteen} we plot the lowest eigenvalue $\sigma ^{2}$ for the solutions lying on that part of the $r_{m}=18$ contour in the $(a_{0}, \phi _{0})$-plane depicted in Fig.~\ref{fig:seven}.
Fig.~\ref{fig:twelve} shows $\sigma ^{2}$ as a function of the parameters $a_{0}$ and $\phi _{0}$ on this contour, while Fig.~\ref{fig:thirteen} shows the same data as a function of the soliton electric charge $Q$ (\ref{eq:charge}).
From Fig.~\ref{fig:twelve} we see that the solitons with smaller values of $a_{0}$ and $\phi _{0}$ are stable and have $\sigma ^{2}>0$, while $\sigma ^{2}<0$ and the solutions become unstable for large $a_{0}$ and $\phi _{0}$ on the $r_{m}=18$ contour.
As we have already seen in Fig.~\ref{fig:nine}, the two parts of the $r_{m}=18$ contour correspond to two branches of solutions, one (the ``low charge'' branch) having smaller soliton electric charge $Q$ and the other (the ``high charge'' branch) having larger values of $Q$.
In Figs.~\ref{fig:twelve} and \ref{fig:thirteen} we find both stable and unstable solitons on both branches.
From Fig.~\ref{fig:thirteen}, on the ``low charge'' branch, solitons with smaller values of $Q$ are stable while those with larger $Q$ are unstable. In contrast, for the ``high charge'' branch, it is those solitons with smaller values of $Q$ which are unstable while those with large $Q$ are stable.
We investigated other constant $r_{m}$ contours and found similar behaviour.

\section{Discussion}
\label{sec:conc}

In this paper we have presented new regular soliton solutions of the Einstein-charged scalar field equations in a cavity.  The static, spherically symmetric solutions are regular everywhere inside and on a reflecting boundary at $r=r_{m}$, on which the scalar field vanishes.
As with the corresponding black hole solutions \cite{Dolan:2015dha}, these solitons do not exist in asymptotically flat spacetime in the absence of the mirror-like boundary.

The mirror is placed at the zero of the equilibrium scalar field nearest the regular origin.  The static field equations possess a scaling symmetry which means that we can fix the scalar field charge $q$ without loss of generality.  The system then has a single length scale, set by the radius of the mirror $r_{m}$.
This is in contrast to the black hole case, where there are two length scales: the radius of the mirror $r_{m}$ and the radius of the event horizon $r_{h}$.
The phase space of soliton solutions is therefore simpler than the black hole phase space described in \cite{Dolan:2015dha}.
With the scalar field charge $q$ fixed, the soliton solutions are parameterized by two quantities: the value of the scalar field at the origin $\phi _{0}$, and the electromagnetic potential at the origin, $a_{0}$.

In the black hole case, there is an upper bound on the corresponding phase space parameter describing the electromagnetic field, which arises from the requirement of a regular event horizon.  In the soliton case, there are no {\it {a priori}} constraints on
the parameters $a_{0}$ and $\phi _{0}$.
For each value of $a_{0}$, we find regular soliton solutions in a finite range of values of $\phi _{0}$.  However, we have not been able to find an upper bound on the value of $a_{0}$ for which there are nontrivial soliton solutions.
When $a_{0}$ is very large, the size of the interval in $\phi _{0}$ for which there are soliton solutions increases and the mirror radius can be extremely small.

We then examined the stability of the above soliton solutions under linear, spherically symmetric, perturbations of the metric, electromagnetic potential and scalar field, considering time-periodic perturbations with frequency $\sigma $.
All the solutions we examined with sufficiently large mirror radius $r_{m}$ are such that the lowest value of $\sigma ^{2}$ found is positive (and hence the frequency $\sigma $ is real).
Therefore the solitons appear to be stable if the mirror radius is large.
However, for sufficiently small values of the mirror radius $r_{m}$, corresponding to sufficiently large values of $a_{0}$ and $\phi _{0}$,
we find that for some (but not all) solitons the lowest value of $\sigma ^{2}$ is negative, so that the frequency $\sigma $ is imaginary and the solitons are unstable.
Although the stability of the solitons depends in a complicated way on the parameters $a_{0}$ and $\phi _{0}$, we may understand our results heuristically by considering a fixed, but large, $\phi _{0}$.
As the parameter $a_{0}$ increases, the electric field strength also increases, as does the matter energy density at the origin.
It seems to be the case that if $a_{0}$, and hence the matter energy density at the origin, gets too large, then the soliton becomes unstable.

Are there scalar solitons in analogous situations which share the qualitative stability features we find here?  Our spacetime has a time-like boundary, the reflecting mirror, and a natural analogue would be charged-scalar solitons in adS where the boundary of the spacetime is time-like.
The phase space of charged-scalar solitons in four-dimensional adS is extremely rich \cite{Gentle:2011kv} (see \cite{Dias:2011tj} for a similarly comprehensive study of charged-scalar solitons in adS${}_{5}$).
The work in \cite{Gentle:2011kv, Dias:2011tj} considers in depth the $(M,Q)$-phase space of solutions, where $M$ is the mass of the asymptotically adS solitons. It is therefore difficult to draw analogies with our cavity system since we are unable to consistently define a mass for our soliton solutions.

Instead, we consider a more helpful analogy to be boson stars, that is, solitons in models involving a time-dependent complex scalar field with a self-interaction potential but no electromagnetic field (see, for example, \cite{Jetzer:1991jr} for a review of boson stars in asymptotically flat spacetime).
In asymptotically flat spacetime, ground-state boson stars have a scalar field profile which has no zeros. For these boson stars, if the central density is larger than a particular critical value they are unstable; if the central density is smaller than the critical value the boson stars are stable \cite{Gleiser:1988ih, Lee:1988av, Kusmartsev:1990cr}.
Similar behaviour is observed for charged boson stars in asymptotically flat spacetime \cite{Jetzer:1989us} and also for boson stars in asymptotically adS spacetime \cite{Astefanesei:2003qy}.
A nonlinear analysis \cite{Seidel:1990jh} reveals that an unstable ground-state boson star in asymptotically flat spacetime may collapse to form a black hole or scalar radiation may escape to infinity, with a stable boson star as the end-point configuration.
It is also possible for an unstable boson star to dissipate completely, so that ultimately the spacetime is pure Minkowski.

What then might be the end-point of the instability we have found for some charged-scalar solitons inside a small cavity? One possibility is that the configuration settles into an alternative (stable) charged-scalar soliton, although the presence of the mirror makes this unlikely in our view, as there is no mechanism in this scenario for scalar radiation (and thus charge) to escape the system. We conjecture instead that an unstable charged-scalar soliton, when perturbed, collapses to form a black hole. This black hole could have charged-scalar hair, or could be a Reissner-Nordstr\"om black hole without scalar hair. Recent results \cite{Dolan:2015dha, Sanchis-Gual:2015lje} suggest that the outcome will depend on the mirror radius. One could start by evolving our linear perturbation equations in the time domain, to verify our frequency-domain analysis in this paper. To determine the ultimate fate of the instability would require an evolution of the full nonlinear system employing techniques from numerical relativity \cite{Sanchis-Gual:2015lje, Bosch:2016vcp, Choptuik:2015mma}.

In summary, our investigation complements recent work \cite{Dolan:2015dha, Sanchis-Gual:2015lje, Bosch:2016vcp} which casts fresh light on the fate of the black hole bomb instability. In the Einstein-charged scalar field system, a consensus has emerged: generically, in both the cavity \cite{Sanchis-Gual:2015lje} and adS \cite{Bosch:2016vcp} contexts, a charged black hole in vacuum can evolve towards a hairy configuration which is stable. Here, we have shown that, as expected, the hairy black holes in a cavity are accompanied by a wider class of solitonic solutions; and, further, that both stable and unstable solitons exist. We have conjectured that the unstable solitons collapse into black holes, though this remains to be investigated. An important open question is whether any conclusions drawn from studying the charged superradiant instability will apply in the (potentially astrophysically-relevant) rotating case, where a class of scalar-hairy four-dimensional Kerr black holes \cite{Herdeiro:2014goa, Herdeiro:2015gia, Herdeiro:2015waa, Chodosh:2015oma} appears to be a plausible candidate for end products of the black hole bomb instability.

\begin{acknowledgments}
The work of SRD and EW is supported by the Lancaster-Manchester-Sheffield Consortium for Fundamental Physics under STFC grant ST/L000520/1.
The work of SRD is also supported by EPSRC grant EP/M025802/1.
EW  thanks the University of Canterbury, Christchurch, New Zealand for an Erskine Visiting Fellowship supporting this work.
\end{acknowledgments}

\appendix
\section{Nonexistence of asymptotically flat gravitating charged-scalar solitons}
\label{sec:appendix}

In this appendix we outline the proof of the nonexistence of asymptotically flat, static, spherically symmetric, charged-scalar solitons.
We essentially follow the argument in \cite{Bekenstein:1971hc}, adapted to soliton rather than black hole solutions, and restricted to spherically symmetric configurations only.

We start with the static scalar field equation (\ref{eq:phiprimeprime}), multiply throughout by $-r^{2} \phi {\sqrt {h}}$ and integrate from $r=0$ to $r=\infty $:
\begin{align}
0 &=
\int _{r=0}^{\infty } \left[ -\phi \frac {d}{dr} \left( r^{2} f{\sqrt {h}} \phi ' \right) - \frac {\left( qA_{0} \right) ^{2} }{f{\sqrt {h}}} r^{2} \phi ^{2}
\right] \, dr
\nonumber \\
&=
\left[ -r^{2}f{\sqrt {h}} \phi \phi ' \right] _{r=0}^{\infty }
\nonumber \\
& \qquad
+ \int _{r=0}^{\infty }
\left[  r^{2} f{\sqrt {h}} \phi '^{2}
-\frac {\left( qA_{0} \right) ^{2} }{f{\sqrt {h}}} r^{2} \phi ^{2}  \right] \, dr,
\label{eq:phiint}
\end{align}
where we have performed an integration by parts.
Similarly, taking the electromagnetic field equation (\ref{eq:Aprimeprime}), multiplying throughout by $-r^{2}A_{0}/f{\sqrt {h}}$ and integrating from
$r=0$ to $r=\infty $ gives
\begin{align}
0 &=
\int _{r=0}^{\infty } \left[ -A_{0} \frac {d}{dr} \left( \frac {r^{2}}{{\sqrt {h}}} A_{0}'\right) + \frac {q^{2}r^{2}}{f{\sqrt {h}}}A_{0}^{2}\phi ^{2}
\right] \, dr
\nonumber \\
&=
\left[ - \frac {r^{2}}{{\sqrt {h}}} A_{0} A_{0}'\right] _{r=0}^{\infty } +
\int _{r=0}^{\infty } \left[ \frac {r^{2}}{{\sqrt {h}}} E^{2} + \frac {q^{2}r^{2}}{f{\sqrt {h}}} A_{0}^{2}\phi ^{2} \right] \, dr
\label{eq:Aint}
\end{align}
where $E=A_{0}'$ and we have integrated by parts.
Since all the field variables $f$, $h$, $\phi $ and $\phi' $ must be finite at the origin, the $r=0$ contribution to the boundary terms in (\ref{eq:phiint}, \ref{eq:Aint}) both vanish.

In order to have an asymptotically flat spacetime, we require that the metric functions $f$ and $h$ have the following behaviour as $r\rightarrow \infty $:
\begin{equation}
f = 1+ O(r^{-\Delta }), \qquad h = 1+O(r^{-\Delta }),
\label{eq:metricasympt}
\end{equation}
for some $\Delta >0$.
In (\ref{eq:metricasympt}), we mean that the largest nonunity term in {\em {one}} of $f$ or $h$ is $O(r^{-\Delta })$; it may be that this is the largest subleading term in both $f$ and $h$, but we do not assume that this is necessarily the case.  It is therefore possible that in (\ref{eq:metricasympt}), one (but not both) of the $O(r^{-\Delta })$ terms should be $o(r^{-\Delta })$ according to the strict definition of this notation.  In the following, any terms written $O(r^{-{\tilde {\Delta }}})$ for some ${\tilde {\Delta }}$ should be interpreted to mean ``no larger than $r^{-{\tilde {\Delta }}}$ as $r\rightarrow \infty $'', whether ${\tilde {\Delta }}$ is positive or negative.

Using (\ref{eq:metricasympt}) as described above, the relevant components of the Ricci tensor have the following behaviour as $r\rightarrow \infty $:
\begin{equation}
R_{tt} \sim O(r^{-\Delta - 2}), \quad
R_{rr} \sim O(r^{-\Delta - 2}), \quad
R_{\theta \theta } \sim O(r^{-\Delta }).
\label{eq:Ricciinf}
\end{equation}
The corresponding components of the trace-reversed stress-energy tensor
\begin{equation}
{\tilde {T}}_{\mu \nu } = T_{\mu \nu } - \frac {1}{2} g_{\mu \nu } T,
\end{equation}
where $T=T_{\mu } ^{\mu }$ is the trace of the stress-energy tensor, must have the same behaviour as (\ref{eq:Ricciinf}) as $r\rightarrow \infty $.
The relevant components of the trace-reversed stress-energy tensor are:
\begin{align}
{\tilde {T}}_{tt} &= \frac {1}{2} fE^{2} + q^{2} A_{0}^{2} \phi ^{2},
\nonumber \\
{\tilde {T}}_{rr} &= -\frac {1}{2fh} E^{2}+ \left( \phi '\right) ^{2},
\nonumber \\
{\tilde {T}}_{\theta \theta } &= \frac {r^{2}E^{2}}{2h}.
\end{align}
Considering $R_{\theta \theta }$, we immediately have $A_{0}'\sim O(r^{-1-\Delta /2})$ as $r\rightarrow \infty $.
Then $R_{rr}$ and $R_{tt}$ give, respectively, that $\phi ' \sim O(r^{-1-\Delta /2})$ and $A_{0}\phi \sim O(r^{-1-\Delta /2})$ as $r\rightarrow \infty $.
The field equations (\ref{eq:Aprimeprime}, \ref{eq:phiprimeprime}) then imply that $A_{0}\sim O(r^{-1})$ and $\phi \sim O(r^{-1})$
as $r\rightarrow \infty $.
Combining these conditions on $A_{0}$, $\phi $ and their derivatives as $r\rightarrow \infty $, we deduce that $A_{0}, \phi \sim O(r^{-{\tilde {\Delta }}})$
as $r\rightarrow \infty $, where ${\tilde {\Delta }} = \max \{ 1, \Delta /2 \} $.
Therefore the boundary terms from $r\rightarrow \infty $ in (\ref{eq:phiint}, \ref{eq:Aint}) vanish.

Now turn to (\ref{eq:Aint}).
On the right-hand-side of the equality we have the sum of two positive terms. The only way this sum can be zero is if both positive terms are individually zero.
Therefore it must be the case that $A_{0}'\equiv 0$ for all $r\in [0,\infty )$, and furthermore that $A_{0}\phi \equiv 0$, so that $A_{0}$ is a constant
everywhere and either $A_{0}$ or $\phi $ vanishes identically.

Substituting $A_{0}\phi \equiv 0$ into (\ref{eq:phiint}) leaves a single positive term on the right-hand-side of the equality, which must vanish.
This means that $\phi '\equiv 0$ for all $r\in [0,\infty )$.

In summary, the only possible asymptotically flat soliton solution of the field equations (\ref{eq:static}) is Minkowski spacetime with $f\equiv 1 \equiv h$ and $\phi $, $A_{0}$ constant.
Since we have shown that $\phi $ and $A_{0}$ both tend to zero as $r\rightarrow \infty $, they must both vanish identically.

The above proof begins with the assumption that neither $\phi $ nor $A_{0}$ vanish identically.
If we have $A_{0}\equiv 0$ but $\phi \neq 0$ as a starting point, then (\ref{eq:Aint}) is trivial and the second term in the integral in (\ref{eq:phiint}) vanishes.
In this case, comparing the $rr$ components of the Ricci tensor and trace-reversed stress-energy tensor reveals that $\phi ' \sim O(r^{-1-\Delta /2})$ as $r\rightarrow \infty $. If $A_{0} \equiv 0$, then the field equations (\ref{eq:static}) depend only on $\phi '$ and not on $\phi $.
We may therefore assume, without loss of generality, that $\phi \rightarrow 0$ as $r\rightarrow \infty $.
As a result of this assumption, the boundary term in (\ref{eq:phiint}) vanishes. The integrand in (\ref{eq:phiint}) is a single positive term which must therefore vanish identically, giving $\phi \equiv 0$.

On the other hand, if we assume that $\phi \equiv 0 $ but $A_{0} \neq 0 $, the system reduces to pure Einstein-Maxwell theory, for which it is well-known (see, for example, \cite{heusler}) that there are no nontrivial soliton solutions.

\end{document}